\pdfoutput=1

\documentclass[11pt,a4paper]{article}
\usepackage{jcappub}

\usepackage{graphicx}
\usepackage{epsfig}
\usepackage{color}
\usepackage{comment}
\usepackage{amsthm}
\usepackage{hyperref}
\usepackage{amssymb}
\usepackage{booktabs}
\usepackage{slashed}
\usepackage{physics}
\usepackage{microtype}
\usepackage{dsfont}

\newcommand{\be}{\begin{equation}}
\newcommand{\ee}{\end{equation}}
\newcommand{\bea}{\begin{eqnarray}}
\newcommand{\eea}{\end{eqnarray}}

\newcommand{\gapp}{\mathrel{\raise.3ex\hbox{$>$}\mkern-14mu
\lower0.6ex\hbox{$\sim$}}}
\newcommand{\lapp}{\mathrel{\raise.3ex\hbox{$<$}\mkern-14mu
\lower0.6ex\hbox{$\sim$}}}
\def\bbox{{\,\lower0.9pt\vbox{\hrule \hbox{\vrule height 0.2 cm
\hskip 0.2 cm \vrule  height 0.2 cm}\hrule}\,}}

\title{The Signals of Doomsday I: False Higgs vacuum decay signatures}

\author[a]{Amartya Sengupta,}
\author[a]{Dejan Stojkovic,}
\author[b,c]{De-Chang Dai}

\affiliation[a]{HEPCOS, Department\, of \,Physics, SUNY\, at\, Buffalo, Buffalo, NY\, 14260-1500, USA}
\affiliation[b]{Department of Physics, National Dong Hwa University, Hualien, Taiwan, Republic of China}
\affiliation[c]{CERCA, Department\, of\, Physics,\, Case\, Western\, Reserve\, University, \,Cleveland, OH\, 44106-7079}

\emailAdd{amartyas@buffalo.edu}
\emailAdd{ds77@buffalo.edu}
\emailAdd{diedachung@gmail.com}

\abstract{
The measured standard model parameters indicate that we might live in a false Higgs vacuum, though with a very long lifetime. However, small black holes can serve as catalysers and significantly speed up the phase transition. In fact, bubbles of true vacuum might already exist in our universe. If the propagation of the bubble walls slows down due to interaction with the surrounding matter and plasma, these signals can reach us before the bubble wall hits us. Using the vacuum mismatch method, we calculate the spectrum of the Higgs particles produced by such a bubble until the terminal velocity is reached. In addition, we show that frictional dissipation at the terminal wall velocity generates a large population of thermally produced Higgs particles, which continues even after the mismatch channel shuts off. Since the Higgs is neutral, a good part of the final decay products (after hadronization, annihilation and decay of unstable particles) will be photons and neutrinos, which will then act as a long-range signature. For the conservative set of parameters used here, the thermal channel produces a macroscopically large burst of high–energy neutrinos and photons from Higgs decays, which could be detectable from sufficiently nearby bubbles with current or upcoming multi–messenger facilities.
}

\begin{document}
\maketitle
\tableofcontents

\section{Introduction}
It is believed that our universe has so far undergone a number of phase transitions at different energy scales, e.g. GUT, electroweak, QCD... Phase transitions usually introduce drastic changes in the structure of the universe, and if they happen late enough, they would almost certainly be fatal to any existing life forms.
Until recently, late-time phase transitions have not been a topic of intensive study (see however \cite{Stojkovic:2007dw,Greenwood:2008qp,Isidori:2001bm,Alonso:2023jsi,Espinosa:2025ejf,Elias-Miro:2011sqh,Elias-Miro:2012eoi,Bentivegna:2017qry,Baker:2021sno,Baker:2021nyl,Shakya:2025qpi,Espinosa:2015qea,Linde:1981zj,Dine:1992wr,Linde:1980tt,Linde:1977mm,Dine:1992vs,Kallosh:2003mt,Kallosh:2003bq,Krive:1976sg,Kofman:1986je,Kawana:2022lba,Ai:2023yce}). A similar late–time analysis of $SU(3)_c$ symmetry breaking in a true–vacuum background has been presented in Ref.~\cite{Sengupta:2025cdm} and an analysis for $U(1)_{\rm EM}$ symmetry breaking will appear soon.

After the Higgs discovery, it became apparent that a scenario similar to the one outlined in \cite{Greenwood:2008qp} (that the Higgs potential might develop instability with our current universe sitting in the unstable false vacuum) might actually be true \cite{Degrassi:2012ry,Espinosa:2007qp,Espinosa:2018eve,Shakya:2025mdh,Ellis:2009tp,Hiller:2024zjp}. This might not be an immediate problem, since the lifetime of our universe in the false vacuum could be many billions of years. However, as shown in \cite{Coleman:1977py,Coleman:1980aw,Berezin:1990qs,Gregory:2013hja,Burda:2015yfa,Burda:2016mou,Hamaide:2023ayu,Canko:2017ebb,Strumia:2023awj}, small primordial black holes can play the role of Higgs vacuum decay catalysers and significantly increase the tunneling probability. Depending on the small black hole mass and the exact parameters in the Higgs potential, the tunneling probability could be high enough to produce bubbles of true vacuum within the present lifetime of the universe. In this context, one should distinguish between the scale at which the Higgs quartic first turns negative and the generally higher field value at which a stabilized true minimum may appear once higher-dimensional operators are included; the latter is the relevant scale for the benchmark considered in this work.

Once a bubble of true vacuum is created, it will expand at a speed close to the speed of light. A bubble wall sweeps through the universe and destroys (or modifies beyond recognition) everything it encounters. If the bubble wall is approaching with the speed of light, then no signal emitted from the wall can reach the observer before the wall, simply because the wall is already moving with the maximal possible speed. However, interaction with surrounding plasma and matter could slow down the wall propagation \cite{Dai:2021boq,Balaji:2020yrx,Konstandin:2010dm,Ai:2024shx}. For example, if the bubble is created one million light years away from us (somewhere between us and Andromeda) and the wall slows by only $1$ km/s, then the signal can reach us about three years before the bubble wall. It is therefore of utmost importance to calculate the characteristics of the possible signals that might come from an approaching bubble.

We investigate here the process in which the Higgs tunneling can leave a signature in the form of produced particles that can potentially reach us before the bubble. In particular, we calculate the spectrum of the Higgs particles created due to the vacuum mismatch inside and outside of the bubble. A natural question is why we focus here on Higgs production, rather than on other possible channels such as direct production of lighter fermion pairs or interactions of ambient particles in the interstellar or intergalactic medium with the bubble wall. Our main reason is that the transition under consideration is itself a Higgs-field transition, so Higgs excitations provide the most direct probe of the changing vacuum background. In that sense, the Higgs sector is the most immediate channel to analyze in a first phenomenological study of false-vacuum decay signatures.

A second reason is that Higgs production provides a comparatively direct bridge between the microscopic bubble dynamics and the long-range messengers relevant for observation. Once produced, the heavy Higgs excitations promptly decay into Standard Model particles, whose subsequent decay chains, hadronization, annihilation, and cascades generate photons and neutrinos. These are precisely the observables of interest in the present work.

We do not assume that Higgs bosons are universally the dominant channel in every physical environment. Other processes, including direct production of additional particle species from the time-dependent vacuum background and interactions of the bubble wall with ambient matter, may also contribute and in some settings could be important. However, a systematic treatment of all such channels would substantially broaden the scope of this paper. Our aim here is therefore more limited: to isolate and study one particularly direct and physically transparent channel associated with the Higgs-sector transition itself. For this reason, Higgs production serves as the natural starting point for the present analysis, while a more complete comparison with other production mechanisms is left for future work.

If there is no friction with the environment, this mechanism will continue indefinitely during the bubble expansion and lead to enormous particle production. Since the bubble walls expand almost with the speed of light, the signal will hit us practically at the same time as the wall itself.  
Therefore, to quantify the amount of particles that can reach us before the bubble wall, we introduce friction between the wall and the environment, which leads to a terminal wall velocity slower than the speed of light. We then integrate particle production up to the moment when the terminal velocity is reached. 

In addition to this vacuum–mismatch emission, the frictional dissipation at the terminal velocity produces a sustained thermal bath of Higgs (and other) particles in the shocked plasma behind the wall. This thermal channel remains active even after the direct mismatch mechanism shuts off, and can exceed the mismatch yield by many orders of magnitude. The Higgs particle is unstable, and it will quickly decay into the standard model particles. Since it is neutral, it will decay into an equal number of charged particles and antiparticles (i.e. quarks  and antiquarks, bosons and antibosons, leptons and antileptons). After hadronization, decay of the unstable particles, and annihilation of particles and antiparticles, we will get mostly photons and neutrinos as the final decay products. It is thus safe to assume that most of the energy invested into the Higgs particle production will eventually end up as photons and neutrinos, which will represent a long range signature of the approaching bubble.

During the expansion of a bubble in a first-order phase transition, particles produced at the wall split into two kinematic populations: those that propagate outward into the false-vacuum exterior, and those that enter the true-vacuum interior. While particles falling inward may subsequently decay according to the true-vacuum symmetries, only a small subset of their light decay products can overtake the relativistically moving wall and re-emerge into the false vacuum. This contribution is therefore negligible. In the present work we focus exclusively on the outward-propagating population—namely, particles that decay in the vicinity of the bubble wall and whose decay products travel toward an observer in the false vacuum—consistently neglecting the suppressed inward-to-outward channel.

\section{The Higgs vacuum decay}
In this section, we review the details of the Higgs vacuum decay in the presence of gravity. The high energy effective Higgs potential has been determined by the two-loop calculations in the standard model (without gravity) as
\begin{equation}
V_{\rm SM} (\phi)=\frac{1}{4} \left( \lambda_* + b\ln^2\frac{\phi}{\phi_*}   \right) \phi^4 .
\end{equation}
 The standard model range of parameters is $-0.01 \lesssim \lambda_{*} \lesssim 0$ , $0.1 M_p \lesssim \phi_*\lesssim M_p$, where, $M_p = \sqrt{\frac{1}{8\pi G}}=2.435\times 10^{18}\,\mathrm{GeV,}$ and $b \sim 10^{-4}$ \cite{Degrassi:2012ry,Burda:2015yfa,Burda:2015isa}. This potential indicates that we live in an unstable vacuum, though with a very long lifetime \cite{Andreassen:2017rzq}. 

As shown in \cite{Burda:2016mou} (for earlier work, see \cite{Greenwood:2008qp}), the inclusion of gravity can significantly modify the dynamics of the electroweak phase transition and drastically shorten the lifetime. We can parameterise the modification by including higher-order operators as
 \begin{equation}\label{hp}
V(\phi)=V_{\rm SM} + \frac{\lambda_6}{6}\frac{\phi^6}{M_p^2} + \ldots
\end{equation}
This potential can have a stable true vacuum and supports a first-order phase transition which happens via the nucleation of bubbles of the new vacuum inside the old vacuum. 
 \begin{figure}[!htbp]
   \centering
\includegraphics[width=12.0cm]{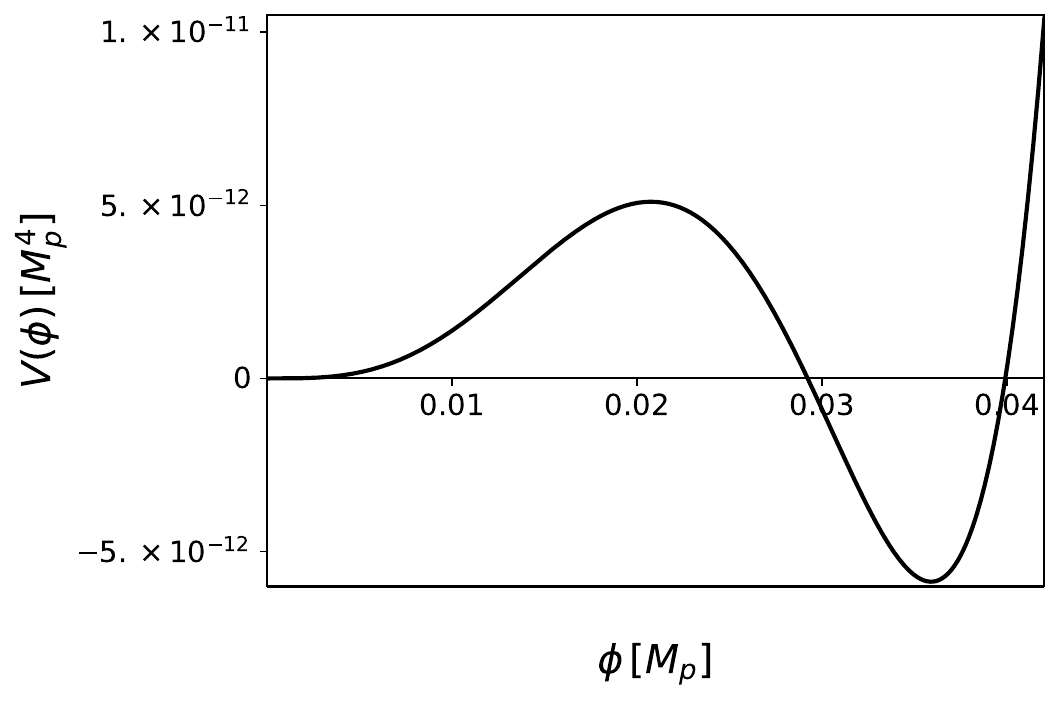}
\caption{This figure shows the Higgs potential $V(\phi)$ in Eq.~(\ref{hp}). The true vacuum is at $\phi\approx 3.6\times 10^{-2}M_p$. The parameters are chosen to be within the standard model, i.e. $b= 10^{-4}$, $\lambda_* = -0.001$, $\phi_* = 0.5M_p$, and $\lambda_6 = 0.34$\cite{Burda:2015isa}.
}
\label{potential}
\end{figure}

We adopt the conventions and follow the calculations in \cite{Burda:2016mou}.
The nucleation rate in the presence of gravity is determined by a bounce solution with Euclidean metric signature (+,+,+,+) with an action
\begin{equation}
S_E = \int_\mathcal{M}\left[-\frac{1}{16\pi G} \mathcal{R}+\left(\frac{1}{2}g^{ab}\partial_a \phi \partial_b \phi +V(\phi) \right)\right]\sqrt{g}d^4x  .
\end{equation}
The spacetime geometry is taken to be spherically symmetric
\begin{equation}
ds^2 = f(r) e^{2\delta(r)} d\tau^2 +\frac{dr^2}{f(r)}+r^2 (d\theta^2 +\sin^2\theta d\varphi^2) ,
\end{equation}
with
\begin{equation}
f(r)=1-\frac{2G\mu (r)}{r} ,
\end{equation}
where $\mu(r)$ is the mass parameter of a black hole remnant.
The phase transition is originally initiated by a black hole of a certain seed mass, $M_+$, related to the black hole remnant mass as
\begin{equation}
M_+=\lim_{r\rightarrow \infty}\mu(r) .
\end{equation}
We distinguish here between the black hole seed mass, $M_+$, which triggers the phase transition, and the black hole remnant mass, $\mu(r)$, which is a leftover from the seed black hole after some of its energy is invested into the phase transition.

The Higgs field equations of motion in this curved background are
\begin{eqnarray}
&&f\phi''+f'\phi'+\frac{2}{r}f\phi'+\delta'f\phi'-V_{,\phi }=0\\
&& \mu'=4\pi r^2 \Big(\frac{1}{2}f{\phi'}^2+V\Big) , \ \ \ \ \delta ' =4\pi Gr{\phi'}^2 ,
\end{eqnarray}
where $V_{,\phi } \equiv \partial V/\partial \phi$. The prime and double prime denote spatial derivatives.
The black hole horizon is at $r=r_h$, which is the solution to $f(r_h)=0$. We will solve these equations numerically in order to get the function $\phi (r)$. To do this, we start from the horizon with a particular remnant parameter, $\mu_-$, at the horizon. The black hole horizon is given as $r_h=2G\mu_-$. At the horizon, the field $\phi$ satisfies the boundary condition
\begin{eqnarray}
&&\mu(r_h)=\mu_- \text{, } \delta(r_h)=0 \\
&&\phi'(r_h)=\frac{r_h V_{,\phi }(\phi(r_h))}{1-8\pi G r_h^2 V(\phi(r_h))} .
\end{eqnarray}
At $r\rightarrow \infty$, the field $\phi$ satisfies $\lim_{r \rightarrow \infty} \phi(r)\rightarrow 0$.
We use a shooting method which starts from a special $\phi$ at $r=r_h$ and check whether $\phi$ becomes $0$ for very large values of $r$. If this condition is satisfied, we have a good solution to the equations of motion.

 \begin{figure}[!htbp]
   \centering
\includegraphics[width=12.0cm]{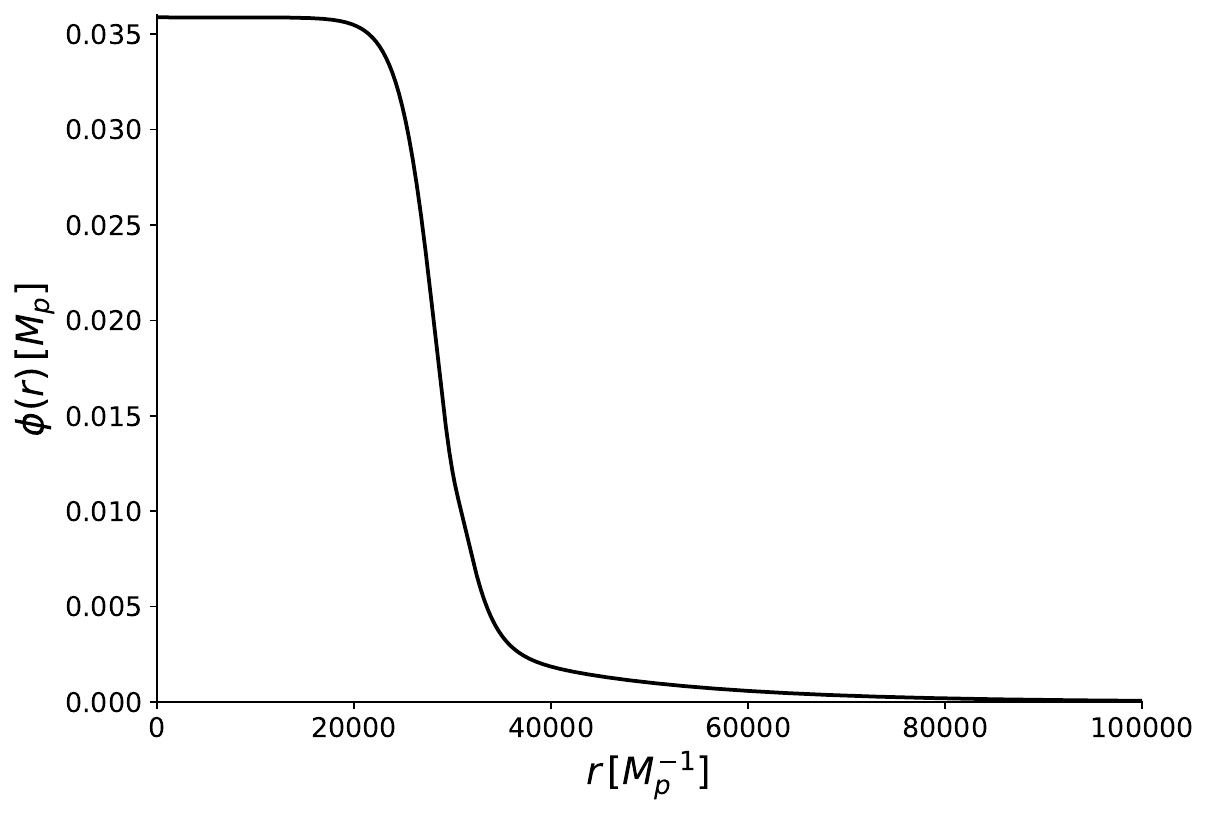}
\caption{ The Higgs field distribution at the moment when a black hole triggers the false vacuum decay. The inner black hole mass is  $1 M_p$. The radius of the true vacuum bubble at that moment is $2\times 10^4 M_p^{-1}$. The parameters are the same as in Fig.~\ref{potential}.
}
\label{field}
\end{figure}

The Higgs potential $V(\phi)$ from Eq.~(\ref{hp}) is shown in Fig.~\ref{potential}, while the solution for the Higgs field distribution in the presence of gravity in Fig.~\ref{field}. Since we have freedom in choosing the exact values of parameters, we follow \cite{Burda:2015isa}, and for illustration we set $b= 10^{-4}$, $\lambda = -0.001$, $\phi_* = 0.5M_p$, and $\lambda_6 = 0.34$. It is important to distinguish between two different field scales that appear in discussions of Higgs vacuum metastability.  The scale often quoted in the literature, of order $10^{10}\,\mathrm{GeV}$ \cite{Espinosa:2015qea}, refers to the field value at which the renormalization-group improved Higgs quartic coupling first turns negative, signalling the onset of metastability.  In the present work, however, the physically relevant scale is not that crossing point itself, but the location of the stabilized high-field true minimum of the effective potential.  Because we include the higher-dimensional operator in Eq.~(\ref{hp}), the potential develops a true minimum at a substantially larger field value.  It is this larger field value that controls the Higgs mass in the bubble interior, the wall thickness, and the characteristic energies of the particles emitted from the bubble wall.

Our benchmark choice is also tied to the black-hole-catalysed vacuum decay scenario.  In that setup, the bubble nucleation process is seeded by a small primordial black hole, and the stabilized true minimum provides the relevant scale entering the bounce configuration and the subsequent bubble dynamics.  We therefore emphasize that the lower instability scale marks the onset of Higgs metastability, whereas the larger fiducial field value used throughout this paper corresponds to the location of the stabilized true vacuum employed in our benchmark analysis.

\section{The true Higgs vacuum bubble propagation}

As shown in \cite{Gregory:2013hja,Burda:2015yfa,Burda:2016mou}, any primordial black hole lighter than $4 \times 10^{14}$g at the time of formation would have evaporated by now, and in the absence of new physics beyond the standard model, would have entered the mass range in which it can trigger false vacuum decay (for related work see \cite{Dai:2019eei,Strumia:2023awj, Chitishvili:2021jrx,Canko:2017ebb,PhysRevLett.111.241801,Branchina:2015nda,Branchina:2016bws,Branchina:2019tyy,Branchina:2018xdh,Bentivegna:2017qry,Espinosa:2025ejf}).
When the Higgs field tunnels through the vacuum barrier, a bubble of true vacuum is created and starts expanding. We assume that the geometry around the bubble is not far from a flat spacetime, so that we can neglect gravitational effects.  In this case, the pure Higgs field, $\phi (\rho)$, action can be written as
\begin{equation}
S(\phi ) =\int d^4x \Big(\frac{1}{2}(\partial_\mu \phi )^2-V(\phi)\Big) ,
\end{equation}
where we neglected other fields that couple to the Higgs. The equation of motion is
\begin{equation}
-\partial_t^2 \phi +\nabla^2\phi- V'(\phi)=0 .
\end{equation}
To proceed, we perform a Wick rotation, $t \rightarrow i\tau$, and the equation of motion becomes
\begin{equation}
\partial_\tau^2 \phi +\nabla^2\phi- V'(\phi)=0 .
\end{equation}
If we consider an $O(4)$ symmetric solution, this equation can be written as
\begin{equation}
\frac{d^2\phi_c}{d\rho^2}+\frac{3}{\rho}\frac{d\phi_c}{d\rho}=V'(\phi_c) .
\end{equation}
In this case, $\rho = \sqrt{\tau^2 +r^2}$, while $\phi_c(\rho)$ depends only on $\rho$. This equation provides a bounce solution with $\partial_t \phi(t=0)=0$ and $\phi(t=0,\vec{x})=\phi_c (\tau=0,\vec{x})$ as
\begin{equation}
\phi(t,\vec{x})=\phi_c (\rho=\sqrt{r^2-t^2}) .
\end{equation}
If we plot $r^2 -t^2=\rho^2=$constant, we will find that the minimal $r$ happens at $t=0$. This is the moment when the bubble bounces back.
Thus, the Higgs bubble is collapsing for $t<0$, bounces at $t=0$, and is expanding for $t>0$. The region $r>t$ is still in the false vacuum, while the $r<t$ region is in the true vacuum. So in this parametrization, nucleation happens at $t=0$, and then the true vacuum bubble expands. We simplify the problem by applying the thin wall approximation\footnote{We model the wall as thin where $\ell_{\rm wall}$ denotes the physical thickness of the bubble wall, i.e. the characteristic distance over which the Higgs background interpolates from the false to the true vacuum. Parametrically this is set by the inverse mass of small fluctuations about the true vacuum,
$\ell_{\rm wall}\sim [V''(\phi_{\rm true})]^{-1/2}\equiv 1/\mu$.
For our benchmark $\mu=7.16\times10^{-4}M_p$, so $\ell_{\rm wall}\simeq 1/\mu \simeq 1.40\times10^{3}\,M_p^{-1}$.
Comparing with the nucleation radius $R_0=2.0\times10^{4}\,M_p^{-1}$ gives the thin-wall parameter
$\varepsilon\equiv \ell_{\rm wall}/R_0 \simeq (1.40\times10^{3})/(2.0\times10^{4})\simeq 0.07$,
so curvature variation across the wall thickness is modest.}
\begin{equation}
\label{thin}
\phi_c(\rho ) = \left\{
  \begin{array}{lr}
    v &  \text{, for $\rho>R$}\\
    v_1 & \text{, for $\rho<R$}
  \end{array}
\right. ,
\end{equation}
where $v$ and $v_1$ are the expectation values of the Higgs field in the false and true vacuum respectively, while $R$ is the radius of the bubble.

Creation of particles during the first-order phase transitions has been studied previously (e.g., \cite{Yamamoto:1994te,MersiniHoughton:1999tt,Mersini-Houghton:1999aoa,Maziashvili:2003kp,Maziashvili:2003sk,Maziashvili:2003zy,Vachaspati:1991tq,Swanson:1986hx,Hamazaki:1995dy,Maziashvili:2003kj,Kobzarev:1974cp,Kawana:2022lba,Espinosa:2010hh,Espinosa:2023oml,Shakya:2023kjf,Mansour:2023fwj}). In principle, particles can be produced via many different mechanisms. For our purpose, we will first use the vacuum mismatch method \cite{Maziashvili:2003zy,Tanaka:1993ez}  which can be easily formulated in Minkowski (as opposed to Euclidean) space, then we will use the thermal production mechanism in the later section.

\section{Particle production due to vacuum mismatch}
\label{pp}

The mismatch of vacua is usually accompanied by particle production. In the present work we focus on the Higgs channel because it is the field directly undergoing the vacuum transition, and therefore provides the most immediate probe of the changing vacuum configuration. In our case, the false Higgs vacuum is outside the bubble, while the true vacuum is inside. As the bubble expands through space, the Higgs field vacuum abruptly changes from false to true. Bogoliubov transformations then indicate that this vacuum mismatch will lead to the Higgs particle production. Since the Higgs particle is very heavy in the new vacuum, its decay products will be very energetic. 
For simplicity, we assume that the Higgs field tunnels to the true vacuum state homogeneously. This should be a good enough approximation since particles are produced locally, i.e. the relevant phase transition length scale is smaller than the size of the bubble at virtually any time after the initial bubble nucleation. Here, for the relevant length scale we take the width of the potential barrier in Fig.~\ref{potential} ($\sim [10^{-2} M_p]^{-1}$), which is also the scale of the Higgs expectation value in the true vacuum.

The fluctuations of the Higgs field, $\phi$, around the background, $\phi_c$, can be written as $\phi =\phi_c +h$, where $h$ is the fluctuation which satisfies the equation of motion
\begin{equation}
\label{motion1}
\partial_{\tilde{\tau}}^2 h +\nabla^2 h- V''(\phi_c)h=0 .
\end{equation}
We again considered only the Higgs part of the Lagrangian and neglected other fields that couple to the Higgs. In this case the equation of motion (\ref{motion1}) can be simplified to
\begin{eqnarray} \label{h}
&&\partial_{\tilde{\tau}}^2 h +\nabla^2 h- M^2 h =0 \text{, for $\tilde{\tau}<\tilde{\tau}^*$}\\
&&\partial_{\tilde{\tau}}^2 h +\nabla^2 h-  \mu^2 h =0 \text{, for $\tilde{\tau}>\tilde{\tau}^*$} ,
\end{eqnarray}
where $\tilde{\tau}^*$ is the characteristic time scale for the duration of the transition between the vacua. 
Since the number density of the produced particles crucially depends on this constant (see Eqs. (\ref{AB}) and (\ref{Nk})) we have to choose it in a meaningful way. We will set $\tilde{\tau}^* = -R_0$, where $R_0$ is the radius of the bubble at the time of nucleation. The justification for this choice is that this is the magnitude of the proper acceleration of the shell, i.e. $a= 1/R_0$. The process of particle creation due to the vacuum mismatch is in this context closely related to Unruh radiation, which is determined by the magnitude of the proper acceleration. We note here that this magnitude is a Lorentz invariant quantity, $a=\sqrt{a_\mu a^\mu}$, which can be calculated from the equation of the bubble motion, $r^2-t^2 = R_0^2$\,(see Appendix-A\,[\ref{App-A}]). As such, it is not equal to the coordinate acceleration that an outside observer would see. The proper acceleration is rather parametrized in terms of the proper time for an observer who is momentarily at rest with respect to the bubble wall. Since in our case $a= 1/R_0$, wall expansion is a motion with a constant proper acceleration, and particles are produced continuously.
Alternatively, we could take $\tilde{\tau}^*=-\kappa^{-1}$, where $\kappa \sim 10^{-2} M_p$ is the characteristic energy scale for the phase transition (i.e. the Higgs vacuum expectation value in the new vacuum, and also the width of the potential barrier). This choice will yield significantly more produced particles, however it lacks connection with the Unruh effect. We therefore go with the more conservative choice $\tilde{\tau}^{*} = -R_0$.

The solution for $h$ in Eq. (\ref{h}) can be written as a combination of mode functions $g_k$, which satisfy $\nabla^2 g_k=-k^2$
\begin{eqnarray}
g_k=\left\{
  \begin{array}{lr}
   e^{\omega_- \tilde{\tau}} e^{i\vec{k}\cdot \vec{x}} &  \text{, for $\tilde{\tau}<\tilde{\tau}^*$}\\
     A_k e^{\omega_+ \tilde{\tau}} e^{i\vec{k}\cdot \vec{x}} +B_k e^{-\omega_+ \tilde{\tau}} e^{i\vec{k}\cdot \vec{x}} & \text{, for $\tilde{\tau}>\tilde{\tau}^*$}
  \end{array}
\right.
\end{eqnarray}
Here, $\omega_{+}=\sqrt{\mu^2+k^2}$ and $\omega_{-}=\sqrt{M^2+k^2}$, while $M=\sqrt{V''(v)}$ and  $\mu=\sqrt{V''(v_1)}$ are the masses of the Higgs field in the false and true vacuum regions, respectively. Since $g_k$ and $\partial_{\tilde{\tau}} g_k$ must be continuous at $\tilde{\tau} =\tilde{\tau}^{*}=-R_{0}$, for $A_k$ and $B_k$ we get
\begin{eqnarray}\label{AB}
A_k&=&\frac{1}{2\omega_+}(\omega_++\omega_-)e^{-(\omega_+-\omega_-)\tilde{\tau}^*}\\
B_k&=&\frac{1}{2\omega_+}(\omega_+-\omega_-)e^{(\omega_++\omega_-)\tilde{\tau}^*} .
\end{eqnarray}

The particle creation spectrum is obtained from the Bogoliubov transform \cite{Tanaka:1993ez}
\begin{eqnarray}
\label{Nk}
N_k=\frac{B_k^2}{A_k^2-B_k^2} =  \left[ \frac{(\omega_++\omega_-)^2}{(\omega_+-\omega_-)^2}e^{4\omega_+ R_0}- 1 \right]^{-1}
\end{eqnarray}
 \begin{figure}[!htbp]
   \centering
\includegraphics[width=10.5cm]{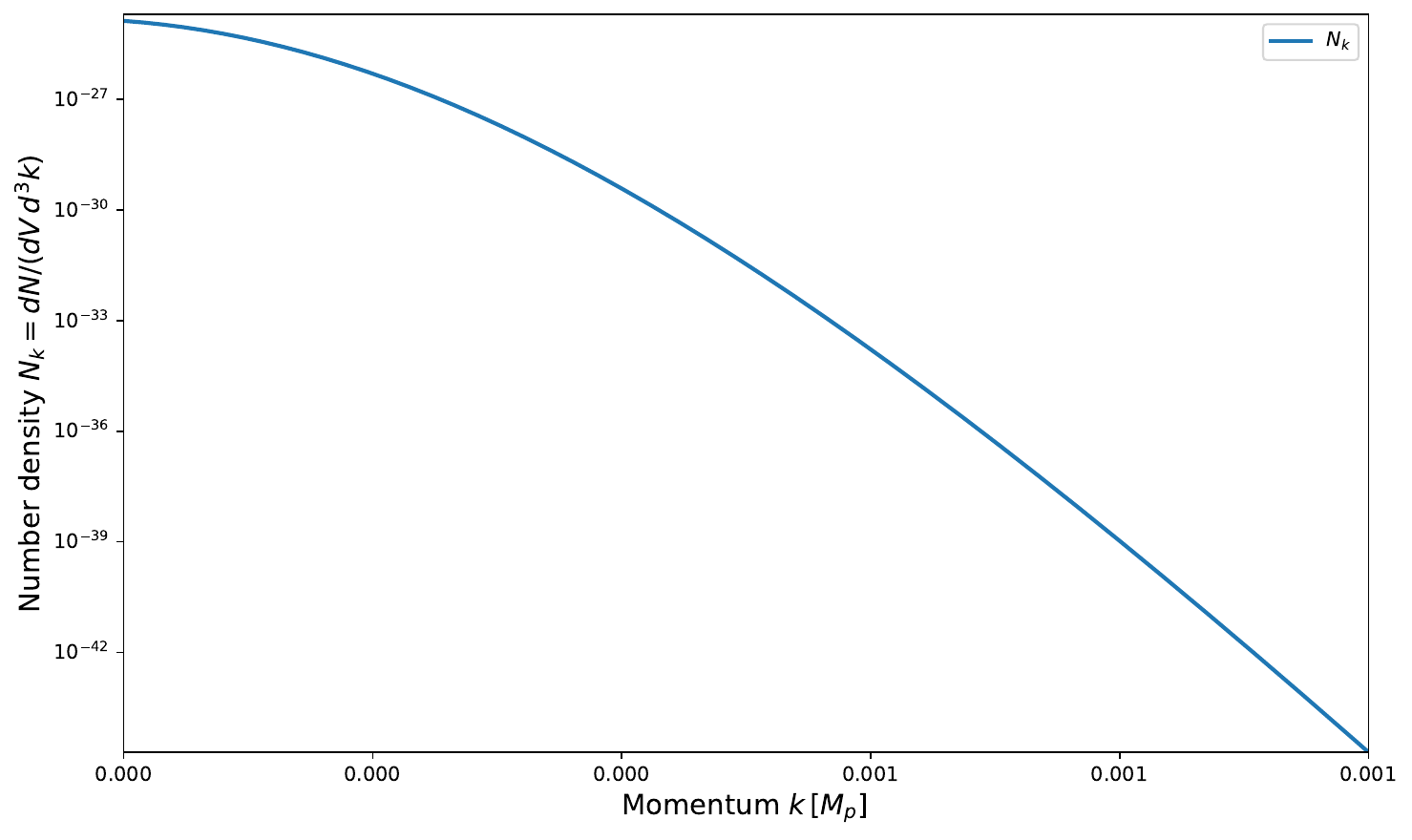} 
\caption{Number density of Higgs particles as a function of their momenta created due to the vacuum mismatch in the Higgs vacuum decay. The units are given in terms of the Planck mass $M_p$.}
\label{nk}
\end{figure}

Fig.~\ref{nk} shows the number density of the produced Higgs particles per momentum mode per unit volume, $N_k = dN/(dV d^3\vec{k})$, as a function of their momenta, $k$, given in Planck units. The value of the Higgs' mass in the false vacuum (where we currently live) is $M =125.09  \,\mathrm{GeV}$ \cite{Aad:2015zhl}. For the choice of parameters we used in Fig.~\ref{potential}, the Higgs mass in the true vacuum is $\mu=7.16\times 10^{-4}M_p$. We emphasize that we do not have much freedom here, and this value might change by only one order of magnitude if we want to keep the values for $b, \lambda$, and $\phi_*$ within the standard model. For the same choice of parameters the bubble radius at the time of nucleation is $R_0 = 2 \times 10^4 M_p^{-1}$.

We can estimate the energy density of Higgs particles inside the bubble (in the new vacuum) at the time of creation as
\begin{eqnarray}
\mathcal{E} &=& \int_0^\infty \omega_+ N_k \frac{d^3\vec{k}}{(2\pi)^3} \nonumber \\
&=& \int_0^\infty \frac{\sqrt{\mu^2+k^2}}{\frac{(\sqrt{\mu^2+k^2}+\sqrt{M^2+k^2})^2}{(\sqrt{\mu^2+k^2}-\sqrt{M^2+k^2})^2} e^{4\sqrt{\mu^2+k^2} R_0} - 1} \frac{d^3\vec{k}}{(2\pi)^3}\nonumber \\
&=& 2.4 \times 10^{-42} M_p^4 = 8.6 \times 10^{31} \, \mathrm{GeV}^4.
\end{eqnarray}

For any bubble of macroscopic volume, this is a very significant total energy which is ultimately transferred into the decay products.  Thus, the bubble of the true vacuum should be a very shiny object. 

A complete treatment of the Higgs decay is very complicated under these circumstances. Inside the bubble the Higgs field is in a new vacuum, so it is heavy, while outside is in the old vacuum, and it is light. A heavy Higgs cannot exist outside of the bubble, and it has to decay practically instantaneously. The decay products have to propagate outside in the old vacuum to reach a distant observer. Strictly speaking, one would have to model how the heavy Higgs propagates through the barrier and decays along the way. To simplify the task, we just assume that the heavy Higgs decays instantaneously at the bubble wall. 

During the expansion of the bubble, vacuum is continuously changing from an old one to a new one, so particles are produced continuously in this process. Since the plot in Figure \ref{nk} is actually particle number density given in Planck units, to get the total number of produced particles (as a function of energy) we have to integrate over the volume of the bubble. We also note that the spectra in Figure \ref{nk} show the particle number density at the site of creation. To obtain the observed spectra on Earth, we have to propagate these particles through expanding space. The corresponding particle flux (number of particles per unit area per unit time) would therefore be diluted by a factor of $1/[4 \pi d^2 (1+z)]$, where $d$ is the physical distance between the source and the observer, while $z$ is the redshift of the source. The factor $(1 + z)$ appears due to the relativistic time delay because particles hit the sphere with the radius $d$ less frequently (two particles emitted $\Delta t$ apart will be measured $(1 + z) \Delta t$ apart. Accordingly, we have to divide the particle energies by a factor of $1+z$ because the individual particle's energies are redshifted by that factor.

\section{Relativistic bubble-wall dynamics in a viscous medium and terminal velocity}
\label{fr}
If the bubble wall moves with the speed of light, then no signal can overtake it to warn us about the impending doom. However, a bubble interacts with its environment and can reach a terminal velocity that is less than the speed of light. Previously, this topic has been studied extensively in the literature, with works addressing bubble wall velocities, hydrodynamical constraints, and friction mechanisms in various contexts\cite{Moore_2000, Moore:1995ua, Branchina:2025jou, PhysRevD.108.103523, Gouttenoire:2021kjv, Bodeker:2017cim, Megevand:2009gh, Kubota:2025avm,Azatov:2020ufh,Branchina:2025adj,Wang:2020zlf,Ramsey-Musolf:2025jyk,Laurent:2022jrs,Si:2025vdt,Dorsch:2018pat}. We begin by noting that a bubble wall propagating through matter or radiation (where some of it could be produced by the bubble itself) experiences friction due to particle interactions at its surface. Understanding this damping is crucial, as friction reduces the wall’s acceleration and sets its terminal velocity, thereby shaping the overall dynamics and determining the efficiency of associated particle production.

\subsection{Derivation of the equation of motion}

In this section, we present a step-by-step derivation of the relativistic equation of motion for a \emph{spherical} thin bubble wall, incorporating the driving vacuum pressure, the Laplace (curvature) pressure, and a linear frictional damping that arises from interactions with the surrounding particles.\\
\medskip
The energy per unit area of the static wall is its surface tension,
\begin{equation}
E_{\rm rest}=\sigma,
\label{eq:rest_energy}
\end{equation}
which under a Lorentz boost becomes
\begin{equation}
E_{\rm wall}=\sigma\,\gamma(v),
\qquad
\gamma(v)=\frac{1}{\sqrt{1-v^2}},
\label{eq:boosted_energy}
\end{equation}
while the corresponding momentum per unit area is
\begin{equation}
\Pi_{\rm wall}=E_{\rm wall}\,v=\sigma\,\gamma(v)\,v.
\label{eq:wall_momentum}
\end{equation}
Here, $v$ is the wall velocity.
Taking the time derivative of equation~\eqref{eq:wall_momentum} yields the inertial response of the wall:
\begin{equation}
\frac{d}{dt}\Pi_{\rm wall}
=\sigma\,\frac{d}{dt}[\gamma(v)\,v]
=\sigma\,\gamma^3(v)\,\frac{dv}{dt},
\label{eq:momentum_derivative}
\end{equation}
where we have used
\begin{equation}
\frac{d}{dt}[\gamma v]
=\bigl[\gamma + v\,\gamma'(v)\bigr]\,\frac{dv}{dt}
=\gamma^3(v)\,\frac{dv}{dt}.
\label{eq:gamma_derivative}
\end{equation}
This result encapsulates the relativistic increase in inertia as $v\to1$.

Following the standard phenomenological treatment of relativistic bubble-wall damping, we model the frictional pressure entering Eq.\eqref{eq:wall_EOM} by the linear-drag form $P_{\rm fric}=\eta\,\gamma(v)\,v$ (see e.g. \cite{Long:2024sqg}).  Ultrarelativistic microphysical calculations allow for different $\gamma$ -scalings of the thermal pressure, $P_{\rm th}\propto \gamma^{0,1,2},$ depending on the assumed kinematics and emission channels; we defer a detailed discussion of this scaling and its implications for our parametrization to Sec.\ref{Scaling_of_the_drag}.

\medskip
For a spherical wall of radius $R(t)$, the net \emph{outward} force per unit area combines the vacuum-pressure jump, the Laplace pressure, and the frictional pressure:
\begin{equation}
F_{\rm net}
\;=\;
\underbrace{\Delta V}_{\text{vacuum drive}}
\;-\;
\underbrace{\frac{2\sigma}{R(t)}}_{\text{Laplace curvature}}
\;-\;
\underbrace{\eta\,\gamma(v)\,v}_{\text{linear drag}}\,.
\label{eq:wall_EOM}
\end{equation}
where $\Delta V = V_{\text{false}} - V_{\text{true}}$ is the pressure difference (latent heat), while $\eta$ is the friction coefficient.
Balancing momentum change against the net force gives the spherical thin-wall equation of motion (EOM):
\begin{equation}
\sigma\,\gamma^3(v)\,\frac{dv}{dt}
= \Delta V \;-\; \frac{2\sigma}{R(t)} \;-\; \eta\,\gamma(v)\,v.
\label{eq:wall_EOM_spherical}
\end{equation}
Equation~\eqref{eq:wall_EOM_spherical} is the relativistic friction-damped \emph{spherical} wall equation. It shows that as the wall accelerates, both the relativistic inertia (via $\gamma^3$) and the frictional drag (via $\gamma v$) increase, while curvature adds an extra restoring pressure $\sim 2\sigma/R$ that is important at early times (small $R$) and fades at late times ($R\to\infty$).

At the instant of terminal balance, the acceleration vanishes ($dv/dt=0$). Equation~\eqref{eq:wall_EOM_spherical} then gives the curvature-modified balance condition
\begin{equation}
0
= \Delta V \;-\; \frac{2\sigma}{R} \;-\; \eta\,\gamma(v)\,v
\quad\Longleftrightarrow\quad
\Delta V_{\rm eff}(R)\;\equiv\;\Delta V-\frac{2\sigma}{R}
= \eta\,\gamma(v)\,v.
\label{eq:balance_spherical_exact}
\end{equation}
In the late-time, large-radius limit $R\to\infty$ one recovers the planar terminal relation $\Delta V=\eta\,\gamma v$.

\subsection{Terminal Velocity}
Calculating the terminal velocity of the wall explicitly requires the knowledge of the frictional coefficient, $\eta$, in a given medium, which is possible in some simple situations, but not in the general case.

The bubble wall reaches terminal velocity when the driving force balances the friction force. The equation of motion for the wall (neglecting its curvature) is:
\begin{equation}
\Delta V = \eta \gamma v.
\end{equation}
For highly relativistic walls ($\gamma \gg 1$, $v \approx 1$), this simplifies to:
\begin{equation}
\Delta V \approx \eta \gamma.
\end{equation}
Thus, the terminal Lorentz factor is:
\begin{equation}
\gamma_{\text{term}} \approx \frac{\Delta V}{\eta}.
\end{equation}
The terminal velocity is then:
\begin{equation}
v_{\text{term}} \approx 1 - \frac{1}{2\gamma_{\text{term}}^2} .
\end{equation}
The friction coefficient $\eta$ can be calculated from microphysics. For a scalar field $\phi$ coupled to other particles (e.g., fermions) with mass $m \sim g \phi$, 
the friction coefficient produced by a relativistic plasma at temperature $T$ can be estimated 
on dimensional grounds as
\begin{equation}
\eta \sim g^2 T^4,
\end{equation}
where $g$ is an effective coupling constant to the thermal bath. 

In the late universe, we have more complicated situations where the bubble is created in the late universe in vacuum, and encounters stars, planets, and diluted interstellar and intergalactic gas during its expansion, more detailed numerical simulations are warranted. After all, the bubble produces a large number of very high energy particles, so it is literally engulfed in its own plasma. Thus, precise calculations of the terminal velocity will be studied elsewhere, and we will instead consider several conservative values for illustration.  

\subsection{Proper acceleration of a spherical bubble wall in a viscous medium}
\label{sec:proper-accel-spherical}
Since particle production we are describing here is a variant of the Unruh effect \cite{Good:2013lca, Walker:1984vj} driven by the proper acceleration of the wall, we will derive now the invariant (proper) acceleration of a relativistic spherical thin wall when both curvature and linear friction are present. The result provides the quantity that directly controls vacuum-mismatch particle production.

We start from the bubble wall equation of motion given in Eq.~(\ref{eq:wall_EOM_spherical})

\begin{equation}
\sigma\,\gamma^3(v)\,\frac{dv}{dt}
= \Delta V - \frac{2\sigma}{R(t)} - \eta\,\gamma(v)\,v.
\label{eq:sph-lab}
\end{equation}
We introduce the wall’s \emph{proper time} $\tau$ and the rapidity $y(\tau)$ via
\begin{equation}
v=\tanh y,\qquad
\gamma=\cosh y,\qquad
\gamma v=\sinh y,\qquad
\frac{dt}{d\tau}=\gamma,\qquad
\frac{dR}{d\tau}=\gamma v=\sinh y.
\label{eq:kinematics}
\end{equation}
The invariant (proper) acceleration is defined by
\begin{equation}
\alpha(\tau)\;\equiv\;\frac{dv}{d\tau}\,\gamma^2
\;=\;\gamma^3\,\frac{dv}{dt}
\;=\;\frac{dy}{d\tau}.
\label{eq:alpha-def}
\end{equation}
Dividing \eqref{eq:sph-lab} by $\sigma$ and using \eqref{eq:alpha-def} yields the proper-time form:
\begin{align}
\frac{dy}{d\tau} &= \underbrace{\frac{\Delta V}{\sigma}}_{\displaystyle A}
\;-\;\underbrace{\frac{2}{R(\tau)}}_{\text{curvature}}
\;-\;\underbrace{\frac{\eta}{\sigma}}_{\displaystyle B}\,\sinh y,
\label{eq:dy-dtau}\\[0.25em]
\frac{dt}{d\tau}&=\cosh y,\qquad
\frac{dR}{d\tau}=\sinh y.
\label{eq:dtR-dtau}
\end{align}
Equivalently, the proper acceleration is \emph{explicitly}
\begin{equation}
{\;
\alpha(\tau)
= A\;-\;\frac{2}{R(\tau)}\;-\;B\,\sinh y(\tau)\,,
\qquad
A\equiv\frac{\Delta V}{\sigma},\quad
B\equiv\frac{\eta}{\sigma}\; }.
\label{eq:alpha-explicit}
\end{equation}\\
We take the nucleation initial data
\begin{equation}
R(0)=R_0,\qquad y(0)=0\ (v=0),\qquad t(0)=0.
\label{eq:ICs}
\end{equation}
From \eqref{eq:alpha-explicit} the \emph{initial} proper acceleration is
\begin{equation}
\alpha(0)=A-\frac{2}{R_0}.
\label{eq:alpha0}
\end{equation}
In the thin-wall nucleation one has the standard relation $A\simeq 3/R_0$, yielding
\begin{equation}
\alpha(0)\;\simeq\;\frac{1}{R_0},
\label{eq:a0}
\end{equation}
which is just the acceleration $a$ we used in Section \ref{pp}. 

For $\tau\ll R_0$ the motion is mildly relativistic. Using $y(\tau)=\alpha(0)\tau+\mathcal O(\tau^2)$, we have for small $y$ $\sinh y\simeq y,$ hence
\begin{equation}
\sinh y\simeq \alpha(0)\,\tau,\qquad
R(\tau)\simeq R_0+\int_0^\tau \sinh y\,d\tau'
\;\simeq\;R_0+\frac{\alpha(0)}{2}\,\tau^2.
\label{eq:early-kin}
\end{equation}
Then by inserting \eqref{eq:early-kin} into \eqref{eq:alpha-explicit} we get the friction term:
\begin{equation}
-B\,\sinh y(\tau)\simeq -B\,\alpha(0)\,\tau (\text{linear in}\, \tau).
\end{equation}
To simplify the curvature term we expand $1/R(\tau)$ about $R_0$:
\begin{equation}
\frac{1}{R(\tau)}=\frac{1}{R_0+\frac{\alpha(0)}{2}\tau^2}
=\frac{1}{R_0}\,\frac{1}{1+\frac{\alpha(0)}{2R_0}\tau^2}
\simeq \frac{1}{R_0}\left(1-\frac{\alpha(0)}{2R_0}\tau^2\right),
\end{equation}
so
\begin{equation}
-\frac{2}{R(\tau)}\simeq -\frac{2}{R_0}+\frac{\alpha(0)}{R_0^2}\,\tau^2.
\end{equation}

\begin{equation}
\alpha(\tau)\simeq
\underbrace{\Big(A-\frac{2}{R_0}\Big)}_{\alpha(0)}
\;-\;B\,\alpha(0)\,\tau
\;+\;\frac{\alpha(0)}{R_0^2}\,\tau^2
\;+\;\mathcal O(\tau^3).
\label{eq:alpha-series}
\end{equation}
With the thin–wall nucleation relation $A\simeq 3/R_0$, the initial proper acceleration is
\begin{equation}
\alpha(0)=A-\frac{2}{R_0}\;\simeq\;\frac{1}{R_0}\equiv a.
\label{eq:a0_new}
\end{equation}
Inserting $\alpha(0)=1/R_0$ into the early–time series $\alpha(\tau)\simeq \alpha(0)-B\,\alpha(0)\tau+\alpha(0)\tau^2/R_0^2+\mathcal O(\tau^3)$ yields
\begin{equation}
\alpha(\tau)\;\simeq\;
\frac{1}{R_0}\;-\;\frac{B}{R_0}\,\tau\;+\;\frac{\tau^2}{R_0^3}
\;+\;\mathcal O\!\big(\tau^3\big),
\label{eq:alpha-series-simplified}
\end{equation}
valid in the mildly–relativistic regime $\tau\ll R_0$ (so $y\ll1$) and for weak friction in the sense $B R_0\ll1$. In this expansion, the \emph{linear} term $-(B/R_0)\tau$ comes from the viscous drag $-B\,\sinh y$, while the \emph{quadratic} term $+\tau^2/R_0^3$ originates from the curvature relaxation in $-2/R(\tau)$ via $R(\tau)\simeq R_0+\tfrac12 \alpha(0)\tau^2$. Higher–order pieces $\mathcal O(\tau^3)$ include small corrections from the next orders in the $\sinh y$ and $1/R$ expansions and are negligible under the same conditions.

Treating \eqref{eq:alpha-series-simplified} as a quadratic polynomial in $\tau$, $\alpha(\tau)\approx a\,\tau^2+b\,\tau+c$ with $a=1/R_0^3$, $b=-(B/R_0)$ and $c=1/R_0$, this parabola (with $a>0$) minimum occurs at
\begin{equation}
\tau_\star
=\frac{-b}{2a}
=\frac{(B/R_0)}{2/R_0^3}
={\;\frac12\,B\,R_0^2\;},
\qquad
\alpha_{\min}
=c-\frac{b^2}{4a}
={\;\frac{1}{R_0}-\frac14\,B^2\,R_0\;}.
\label{eq:alpha-min}
\end{equation}
$\tau_\star$ is the early-time proper time (measured on the wall) at which the quadratic Taylor approximation for the proper acceleration reaches its local minimum. Physically, $\tau_\star$ marks the instant where curvature’s quadratic ``recovery'' just balances the linear drag trend, and $\alpha_{\min}$ is the corresponding (minute) reduction of the net drive. The impact on instantaneous vacuum–mismatch production is therefore negligible at early times. Beyond this local regime, as $y$ grows the friction term $-B\sinh y$ dominates and the full evolution drives $\alpha(\tau)\to0^+$ toward terminal balance; thus $\alpha_{\min}$ here is an \emph{early–time} feature of the quadratic approximation, not a global minimum of the complete dynamics.


We differentiate \eqref{eq:alpha-explicit} and use \eqref{eq:dtR-dtau}:
\begin{align}
\frac{d\alpha}{d\tau}
&= \frac{d}{d\tau}\!\left(A-\frac{2}{R}-B\sinh y\right)
= \frac{2}{R^2}\,\frac{dR}{d\tau} \;-\; B\,\cosh y\,\frac{dy}{d\tau}
\nonumber\\
&= \frac{2\,\sinh y}{R^2}\;-\;B\,\cosh y\,\alpha.
\label{eq:alpha-prime-exact}
\end{align}
In the planar, small–rapidity limit (\(R\to\infty\), \(\cosh y\simeq1\)) one has \(d\alpha/d\tau\simeq -B\,\alpha\), so \(\alpha(\tau)\simeq \alpha(0)e^{-B\tau}\)\label{eq:alpha-bound}. For spherical walls, the positive curvature term slows this decay; we therefore use \(\tau_{\rm fric}\equiv 1/B=\sigma/\eta\) as a characteristic timescale for approaching terminal balance.
From this we get 
\begin{equation}
\tau_{term}=1/B = \frac{\sigma}{\eta}\label{TerminalTime} ,
\end{equation}\\
which defines the characteristic time scale for the terminal velocity to be reached. We justify the cutoff at $\tau_{\rm term}$ for the vacuum--mismatch channel and the choice $\tau_{\rm final}$ for the thermal channel in Appendix~\ref{App-C}.
As $R\to\infty$ the curvature term disappears and \eqref{eq:alpha-explicit} reduces to
\begin{equation}
\alpha(\tau)\;\to\;A-B\,\sinh y(\tau).
\label{eq:alpha-planar-limit}
\end{equation}
The terminal state is reached when $\alpha\to 0$, i.e.
\begin{equation}
\sinh y_{\rm term}=\frac{A}{B}
\quad\Longleftrightarrow\quad
\gamma_{\rm term}\,v_{\rm term}=\frac{\Delta V}{\eta}.
\label{eq:terminal}
\end{equation}

The quantity $\tau_{\rm term}$ from \eqref{TerminalTime} is important because it marks the onset of the quasi–terminal regime. Vacuum–mismatch particle production is dominant during  the interval $\tau \lesssim \tau_{\rm term}$, while for $\tau \gtrsim \tau_{\rm term}$ the proper acceleration has decayed on this timescale, so $N_{k=0}(\tau)$, and hence $\mathrm{d}N_{k=0}/\mathrm{d}\tau$, become exponentially small. The corresponding numerical bounds, which show that the late–time contribution to integrated particle yield beyond $3\tau_{\rm term}$ or $5\tau_{\rm term}$ is completely negligible, are presented in Appendix~\ref{App-C}. Particles may still be produced at later times via other mechanisms (as we will show below), but not through this Unruh–like vacuum–mismatch channel, which is controlled by a non–zero proper acceleration.

\section{Particle production from vacuum mismatch limited by friction}
\label{sec:higgs-prod-spherical}

With the friction dynamics in hand, we now estimate how many Higgs quanta are produced in the regime before the terminal velocity is reached.  Throughout we work in Planck units, $G=\hbar=c=1$. The relevant parameters taken from our choice of the potential in the previous sections are
\begin{equation}\nonumber
\Delta V = 5.869\times10^{-12}\;M_p^4,
\quad
R_0      = 2.0\times10^{4}\;M_p^{-1},
\quad
\mu      = 7.16\times10^{-4}\;M_p,
\end{equation}
\begin{equation}
\sigma   = 5.868\times10^{-8}\;M_p^3,
\quad
M        = 125.09\,\mathrm{GeV}.
\label{eq:inputs}
\end{equation}
We use again dimensionless drive and friction ratios
\begin{equation}
A\equiv\frac{\Delta V}{\sigma},\qquad
B\equiv\frac{\eta}{\sigma},
\qquad
A=1.001\times10^{-4}\;M_p,
\end{equation}
and parametrize the wall by its proper time $\tau$ and rapidity $y(\tau)$:
\begin{equation}
v=\tanh y,\quad
\gamma=\cosh y,\quad
\gamma v=\sinh y,\qquad
\frac{dt}{d\tau}=\gamma,\quad
\frac{dR}{d\tau}=\sinh y.
\label{Parameters}
\end{equation}
The \emph{proper} acceleration that controls particle production is
\begin{equation}
\alpha(\tau)\;\equiv\;\frac{dy}{d\tau}
\;=\;A\;-\;\frac{2}{R(\tau)}\;-\;B\,\sinh y(\tau),
\label{eq:alpha-spherical}
\end{equation}
with nucleation data
\begin{equation}
R(0)=R_0,\qquad y(0)=0\ (v=0),\qquad t(0)=0.
\end{equation}
Note again that $\alpha(0)=A-2/R_0\simeq 1/R_0$ under the thin-wall relation $A\simeq 3/R_0$, and that curvature $2/R$ initially suppresses the drive but becomes negligible as $R$ grows.

Using the proper acceleration $\alpha(\tau)$ from \eqref{eq:alpha-spherical} we explicitly integrate particle production up to $\tau=\tau_{\rm term}$. Since both the terminal velocity, $v_{\rm term}$, and the time scale $\tau_{\rm term}$ depend on the friction coefficient $\eta$, for illustration, we will consider a few values of $v_{\rm term}$ for which we will calculate $\tau_{\rm term}$ and perform numerical integration.  


At proper time $\tau$, we generalize the instantaneous zero-momentum mode occupancy calculated in eq. (\ref{Nk})  using the vacuum mismatch procedure 
\begin{equation}
N_{k=0}(\tau)
=\left[ \frac{(\omega_++\omega_-)^2}{(\omega_+-\omega_-)^2}e^{\frac{4\omega_+}{\alpha (\tau)}}- 1 ,\right]^{-1},
\label{eq:Nk0-alpha}
\end{equation}
where we replaced $R_0 = 1/\alpha (0)$ with $R_0 = 1/\alpha (\tau)$ to take into account that the proper acceleration decreases due to friction. Also, after setting $k=0$, we have $\omega_+ = \mu$ and $\omega_- = M$. This $N_{k=0}(\tau)$  vanishes automatically when $\alpha(\tau)=0$ (terminal balance). During a small proper-time step $d\tau$, the shell of volume swept by the wall is $4\pi R^2\,v\,dt=4\pi R^2\,\sinh y\,d\tau$. Multiplying this new volume by the instantaneous number per unit volume $N_{k=0}(\tau)$ gives the increment in the number of particles produced. Therefore, the instantaneous number production is the occupancy times the area and the lab-time increment:
\begin{equation}
dN
= N_{k=0}(\tau)\;4\pi R(\tau)^2\;v(\tau)\,dt
= N_{k=0}(\tau)\;4\pi R(\tau)^2\;\sinh y(\tau)\,d\tau,
\end{equation}
so that the accumulated yield satisfies
\begin{equation}
\frac{dN_{k=0}}{d\tau}
= N_{k=0}(\tau)\;4\pi R(\tau)^2\;\sinh y(\tau),
\qquad
N^{(\text{int})}_{k=0}(0)=0.
\label{eq:dNtot}
\end{equation}
$N_{k=0}^{\rm (int)}$ is the cumulative number produced up to time $\tau$ by integrating those increments beginning at nucleation. At the instant of nucleation $(\tau=0)$ no time has elapsed and no volume has been swept, so by definition the accumulated total is zero. $N_{k=0}(0)$ is, on the other hand, an instantaneous occupation per unit volume; we still need a finite swept volume (over a finite $d\tau$) to accumulate a non-zero total.\\

\begin{table*}[tbhp]
\centering
\small
\begin{tabular}{@{}lccccc@{}}
\toprule
Scenario & $\eta\,[M_p^4]$ & $\tau_{\rm term}\,[M_p^{-1}]$ & $R_{\rm fin}\,[M_p^{-1}]$ & $N_{k=0}^{\rm (int)}$ \\
\midrule
$\delta=10^{-8}$  &
$8.300\mathrm{e}{-16}$ &
$7.069\mathrm{e}{+07}$ &
$7.208\mathrm{e}{+07}$ &
$1.052\mathrm{e}{+08}$ \\
$\delta=10^{-9}$  &
$2.624\mathrm{e}{-16}$ &
$2.235\mathrm{e}{+08}$ &
$2.320\mathrm{e}{+08}$ &
$3.530\mathrm{e}{+09}$ \\
$\delta=10^{-10}$ &
$8.300\mathrm{e}{-17}$ &
$7.069\mathrm{e}{+08}$ &
$7.147\mathrm{e}{+08}$ &
$1.136\mathrm{e}{+11}$ \\
\bottomrule
\end{tabular}
\caption{For each ultra-relativistic deficit $\delta=1-v_{\rm term}$ we evolve the spherical wall with proper acceleration \eqref{eq:alpha-spherical} and kinematics to the time $\tau_{\rm term}=\sigma/\eta$ where the terminal velocity is reached. The proper acceleration $\alpha(\tau)$ directly feeds the instantaneous occupancy \eqref{eq:Nk0-alpha}; the total yield $N_{k=0}^{\rm (int)}$ follows from \eqref{eq:dNtot}. We note that we took into account only the Higgs particles produced at rest, i.e. the $k=0$ mode. The number of particles produced in all momentum modes is higher than shown here. }
\label{tab:spherical-yields}
\end{table*}

The details of the numerical calculation have been described in Appendix~\ref{App-B}. The results show that as the radius grows, friction balances the drive, with the system approaching terminal velocity on the time-scale $\tau_{\rm term}$; and the integrated yield is dominated by the combination of the exponentially sensitive $N_{k=0}(\alpha)$ and the rapidly growing area factor $4\pi R^2$, which becomes substantial by the time $\tau \sim \tau_{\rm term}$ in the most ultra-relativistic cases. However, the production cut-off time $\tau_{\rm term}$ is very short. The number of particles shown in the Table \ref{tab:spherical-yields} might be sufficient if the bubble is produced near us (after all a few detected particles at enormous energies will represent an intriguing signal), but their flux will be severely diluted when propagating over the cosmological distances. However, all the difference in the energy density between the false and true vacuum has to go somewhere. If friction practically quenched particle production due to the vacuum mismatch, most of that energy will go into thermal particle production because of intense heat absorbed by the environment.  

In this section we intentionally worked with an idealised setup where the drag coefficient $\eta$ is held fixed in time.  In other words, the friction ratio $B=\eta/\sigma$ does not respond to how hot or dense the shocked plasma becomes.  The point of this toy model is to cleanly isolate the ``vacuum–mismatch’’ contribution to particle production: an accelerating bubble wall imposes a time–dependent boundary condition on the Higgs field, and the non–adiabatic change in the mode frequencies $\omega_\pm$ produces particles even if we do not allow the medium to thermalise.  In this sense, the zero–mode occupation number $N_{k=0}(\tau)$ should be interpreted as a minimal, non-thermal lower bound on the Higgs–channel particle yield that follows purely from the acceleration history of the wall.

\section{Thermal particle production from frictional dissipation}
\label{sec:thermal-production}

In this section we will use some of the quantities derived before, but will repeat them anyway for clarity. As the bubble wall propagates through an ambient medium, microscopic interactions between the wall and the surrounding particles exert a frictional pressure
\begin{equation}
P_{\rm fric} = \eta\,\gamma v.
\end{equation}
This friction acts against the vacuum--mismatch force that normally accelerates the bubble wall, thereby reducing the kinetic energy that the wall would otherwise acquire.  
However, the missing kinetic energy is not lost: it is deposited directly into a thin layer of shocked plasma immediately behind the advancing wall.  
Because the wall moves ultra-relativistically, this shocked layer thermalizes essentially instantaneously compared to the timescale on which the wall evolves.  
Our goal is to compute how much thermal energy and, consequently, how many thermal particles are produced by this dissipation mechanism.

The derivation presented here follows a fully self-consistent energy-conservation framework.  
It does not assume any specific microphysical interaction; instead, it relies solely on the difference between the trajectory of a frictionless bubble wall and that of a friction-limited wall.  
This difference encodes all dissipative effects in a completely model-independent way.  
Our treatment is consistent with the relativistic fluid analysis of Landau \& Lifschitz and with the microscopic studies of bubble-wall friction in first-order phase transitions~\cite{Moore_2000, Moore:1995ua, Bodeker:2017cim, Arnold:1993wc,Gouttenoire:2021kjv,Balaji:2020yrx}.  
Thermalization within the shocked layer is treated using standard equilibrium thermodynamics~\cite{Kolb:1990vq,Ai:2021kak}.

In principle, once the wall enters a quasi-steady-state regime, direct scattering and momentum transfer of ambient particles across the moving wall can provide an additional source of particle production, as emphasized in microscopic studies of bubble-wall friction and transition-radiation effects. However, in the present setup we expect this contribution to be subdominant compared with the thermal particle production generated by the much larger frictional energy dissipated into the shocked medium, and we therefore do not consider it here.

Astrophysical environments are highly diverse (interstellar gas, stellar matter, self-generated plasmas).  
The microscopic scattering rates and dominant interaction channels are not known in general.  
To avoid tying our analysis to any specific model, we parametrize all dissipation through the terminal-velocity deficit
\begin{equation}
\delta \equiv 1 - v_{\rm term},
\end{equation}
which measures how close the terminal velocity is to the speed of light.  
The corresponding effective friction coefficient is then fixed by the force-balance condition (in the planar limit):
\begin{equation}
\Delta V = \eta\,\gamma_{\rm term} v_{\rm term}.
\end{equation}
This relation ensures that, for any chosen $\delta$, the wall indeed saturates at $v_{\rm term}$ due to friction which we also showed previously.
Since $\eta(\delta)$ encapsulates all microphysical effects, this framework is fully model-independent.  
We focus on three representative values which we have used throughout the paper as a benchmark,
\[
\delta = 10^{-8}, \qquad 10^{-9}, \qquad 10^{-10},
\]
which capture the deeply ultra-relativistic regime.

\subsection*{Local energy conservation and the correct energy deficit}

The energy stored in the wall per unit area is (as before)
\begin{equation}
E_{\rm wall}=\sigma\,\gamma(t),
\qquad
\gamma(t)=\frac{1}{\sqrt{1-v^2}},
\label{eq:boosted_energy_again}
\end{equation}
where $\sigma$ is the surface tension and $\gamma$ is the Lorentz factor.  This quantity we also used in section \ref{fr}.
Importantly, curvature affects the wall's acceleration, but does \emph{not} change the form of this energy density.\\
\\
To quantify dissipation, we compare two cases:\\
\emph{(i) The frictionless wall}, which accelerates as strongly as the vacuum pressure allows; \\ 
\emph{(ii) The physical wall}, whose acceleration is reduced by friction.\\\\
The frictionless wall obeys
\begin{equation}
\sigma\,\gamma_0^3\frac{dv_0}{dt}
=\Delta V-\frac{2\sigma}{R} ,
\label{eq:frictionless-eom}
\end{equation}
where $v_0$ (and thus $\gamma_0$) corresponds to the wall's velocity without friction, which comes enormously close to the speed of light, but is not exactly equal to it. 
The term $2\sigma/R$ represents the counteracting curvature force, and the remainder provides the net acceleration.\\
As we have already shown in section \ref{fr}, the friction-limited wall satisfies
\begin{equation}
\sigma\,\gamma^3\frac{dv}{dt}
=\Delta V-\frac{2\sigma}{R}-\eta\,\gamma v,
\label{eq:friction-eom}
\end{equation}
where the last term represents the energy lost to the medium per unit time per unit area.\\
The key identity follows by differentiating $\gamma=(1-v^2)^{-1/2}$:
\begin{equation}
\frac{d\gamma}{dt} = \gamma^3 v\,\frac{dv}{dt}.
\label{eq:dgamma-identity}
\end{equation}
This identity ties changes in kinetic energy directly to changes in velocity.  
Applying it to the frictionless trajectory gives
\begin{align}
\frac{d\gamma_0}{dt}
&=\gamma_0^3 v_0 \frac{dv_0}{dt}
=\gamma_0^3 v_0
\left(
\frac{\Delta V - 2\sigma/R}{\sigma \gamma_0^3}
\right)
\\
&\Rightarrow
\frac{d\gamma_0}{dt}
=
\frac{v_0}{\sigma}\left(\Delta V - \frac{2\sigma}{R}\right).\label{eq:EOM-frictionless}
\end{align}
An analogous expression follows for the frictional trajectory:
\begin{equation}
\frac{d\gamma}{dt}
=
\frac{v}{\sigma}
\left(
\Delta V - \frac{2\sigma}{R}
- \eta\gamma v
\right).
\end{equation}
We now define the central object of this analysis:
\begin{equation}
E_{\rm max}(t)=\sigma\,\gamma_0(t),\qquad 
E_{\rm wall}(t)=\sigma\,\gamma(t),
\end{equation}
whose difference
\begin{equation}
\Delta E(t)
=
\sigma\,[\gamma_0(t)-\gamma(t)]
\end{equation}
represents the energy that the wall should have gained in the absence of friction but did not.  
This energy deficit is the true measure of dissipation.

After differentiation we get the master equation:
\begin{equation}
\frac{d}{dt}\Delta E(t)
=
(v_0 - v)\left(\Delta V-\frac{2\sigma}{R}\right)
+ \eta\,\gamma v^2.
\label{eq:dDeltaE-master}
\end{equation}
Both terms on the right hand side vanish when $\eta \to 0$.

\subsection*{Thermalisation of the deficit}

Once energy flows into the medium, it thermalises inside a comoving layer of fixed proper thickness $\ell$, so the total thermal energy is
\begin{equation}
E_{\rm th}(t)
=
\rho_{\rm th}(t)\,A(t)\,\ell,
\qquad
A(t)=4\pi R^2(t).
\end{equation}
Because the thermal energy per unit area is exactly the accumulated deficit, we have
\begin{equation}
\rho_{\rm th}(t)\,\ell = \Delta E(t).
\end{equation}
Differentiating,
\begin{equation}
\frac{d\rho_{\rm th}}{dt}
=
\frac{1}{\ell}
\frac{d}{dt}\Delta E(t),
\end{equation}
and inserting Eq.~\eqref{eq:dDeltaE-master},
\begin{equation}
\frac{d\rho_{\rm th}}{dt}
=
\frac{1}{\ell}
\left[
(v_0-v)\left(\Delta V-\frac{2\sigma}{R}\right)
+ \eta\,\gamma v^2
\right].
\end{equation}
Switching to proper time via $d\tau = dt/\gamma$,
\begin{equation}
\frac{d\rho_{\rm th}}{d\tau}
=
\frac{\gamma}{\ell}
\left[
(v_0-v)\left(\Delta V-\frac{2\sigma}{R}\right)
+ \eta\,\gamma v^2
\right].
\end{equation}
It is convenient to parameterise the proper thickness of the shocked layer in terms of the
same microscopic scale that controls the wall structure itself.  
In a Higgs-like scalar potential, the true--vacuum mass 
$\mu^2 \sim V''(\phi_{\rm true})$ also sets the characteristic width of the bubble wall,
$\delta_{\rm wall} \sim 1/\mu$, i.e.~the distance over which the field interpolates between the
false and true vacua.  
It is therefore natural to assume that the region in which the fluid is strongly disturbed and
thermalised by the wall cannot be parametrically thinner than the wall, and is instead of the
same order.  
We thus take the comoving thickness of the thermalised layer to scale as
\begin{equation}
\ell \;\sim\; \frac{1}{\mu}\,,
\end{equation}
where $\mu$ is the Higgs mass in the true vacuum (and also sets the width of the potential
barrier).  

The heating rate per unit volume is obtained by dividing the energy flux into the layer
by this thickness.  Using the energy--deficit law,
\begin{equation}
\frac{d\rho_{\rm th}}{d\tau}
=
\frac{1}{\ell}
\left[
\gamma (v_0-v)\left(\Delta V-\frac{2\sigma}{R}\right)
+ \eta\,\gamma^2 v^2
\right],
\end{equation}
and substituting $\ell = 1/\mu$ yields the working evolution equation
\begin{equation}
\frac{d\rho_{\rm th}}{d\tau}
=
\mu\,
\gamma (v_0-v)\left(\Delta V-\frac{2\sigma}{R}\right)
+ \mu\,\eta\,\gamma^2 v^2.
\end{equation}

In this form the microscopic input is encoded in the scale $\mu$, which fixes the effective
thickness of the shocked layer and is directly tied to the underlying Higgs potential, while the
macroscopic drag coefficient $\eta(0)$ controls how efficiently the wall motion pumps energy
into this layer. 

The detailed value of $\ell$ is therefore not an independent free parameter: different
$\mathcal{O}(1)$ variations in the proportionality between $\ell$ and $1/\mu$ would only rescale
the overall normalisation of $\rho_{\rm th}$ and $N_{\rm th}$ by an $\mathcal{O}(1)$ factor,
without affecting the qualitative behaviour of the heating or the relative importance of the
two contributions in the brackets.
The two terms now have clear physical meaning:  
(i) the lost--acceleration heating proportional to $(v_0-v)$, and  
(ii) the direct drag heating proportional to $\eta$.  
Both are essential, and both grow dramatically as the wall becomes ultra--relativistic.
\subsection{Thermal particle production}

Once the thermal energy density is known, the temperature follows from equilibrium thermodynamics:
\begin{equation}
T(\tau) =
\left(
\frac{30}{\pi^2 g_*}\,\rho_{\rm th}(\tau)
\right)^{1/4},
\end{equation}
with corresponding equilibrium number density
\begin{equation}
n(\tau)
=
\frac{\zeta(3)}{\pi^2}\,g_*\,T(\tau)^3
=
\frac{\zeta(3)}{\pi^2}
\,g_*^{1/4}
\!\left(\frac{30}{\pi^2}\right)^{3/4}
\!\left[\rho_{\rm th}(\tau)\right]^{3/4}.
\end{equation}
As the wall moves, it sweeps out the comoving volume
\begin{equation}
dV = 4\pi R^2(\tau)\,\sinh y(\tau)\,d\tau,
\end{equation}
so the instantaneous thermal production rate is
\begin{equation}
\frac{dN_{\rm th}}{d\tau}
=
4\pi R(\tau)^2 \sinh y(\tau)\, n(\tau)
=
4\pi R^2(\tau)\sinh y(\tau)\;
\frac{\zeta(3)}{\pi^2} g_*^{1/4}
\left(\frac{30}{\pi^2}\right)^{3/4}
\left[\rho_{\rm th}(\tau)\right]^{3/4}.
\label{eq:dNth_final_v2}
\end{equation}
Therefore thermal production depends entirely on the integrated heating history,
\begin{equation}
\rho_{\rm th}(\tau)
= \int_0^\tau\! d\tau'\;
\mu\left[
\gamma(\tau')\,\big(v_0 (\tau')-v(\tau')\big)\left(\Delta V-\frac{2\sigma}{R(\tau')}\right)
+\eta(\tau')\,
\gamma(\tau')^2v(\tau')^2\right].
\end{equation}
as dictated by the evolution of the energy deficit $\Delta E$. The numerical procedure associated with the thermal production has been discussed in the next subsection but for the details of the general numerical procedure please refer to Appendix~\ref{App-B}

\subsection*{Numerical procedure}
The dynamical system is conveniently summarised as the coupled set
\begin{equation}
\left\{
\frac{dR}{d\tau},\;
\frac{dy}{d\tau},\;
\frac{dy_0}{d\tau},\;
\frac{d\rho_{\rm th}}{d\tau},\;
\frac{dN_{\rm th}}{d\tau}
\right\},
\end{equation}
which is integrated simultaneously from nucleation up to a finite proper time
$\tau_{\rm final}$.  Here $(R(\tau),y(\tau))$ describe the physical wall
subject to friction, while $y_0(\tau)$ is a frictionless reference rapidity
evolved at the same instantaneous radius $R(\tau)$ and governed by
Eq.~\eqref{eq:EOM-frictionless}.  Evolving both systems
in parallel ensures that the energy deficit is always evaluated against the
correct frictionless configuration at the same radius.

Astrophysical microphysics is parametrised by the terminal--velocity deficit
\begin{equation}
\delta \equiv 1 - v_{\rm term},
\end{equation}
which fixes how close the wall comes to the speed of light before friction
balances the vacuum pressure.  For a given $\delta$ one has
\begin{equation}
v_{\rm term} = 1-\delta,
\qquad
\gamma_{\rm term} = \frac{1}{\sqrt{1-v_{\rm term}^2}}.
\end{equation}
At terminal velocity the net force on the wall must vanish, so the outward
vacuum pressure $\Delta V$ is exactly balanced by the drag (in the planar limit) :
\begin{equation}
\Delta V
=
\eta(0)\,\gamma_{\rm term} v_{\rm term},
\end{equation}
which defines a reference drag coefficient
\begin{equation}
\eta(0)
=
\frac{\Delta V}{\gamma_{\rm term} v_{\rm term}}.
\label{eq:eta-target-higgs}
\end{equation}
Smaller $\delta$ corresponds to a more ultra--relativistic wall and hence to
a smaller $\eta(0)$; because $\gamma_{\rm term}$ grows extremely
rapidly as $v_{\rm term}\to 1$, even modest changes in $\delta$ can change
$\eta(0)$ by many orders of magnitude.  This motivates our use of
$\delta$ as the primary input parameter.

In reality the friction is not constant.  As the wall deposits energy into the
surrounding medium, a shocked layer builds up and pushes back on the wall with
a force that grows with the local temperature.  To capture this behaviour we
promote the drag coefficient to a temperature--dependent quantity
\begin{equation}
\eta(\tau)
=
g_{\rm eff}^2\,T_{\rm eff}^4(\tau),
\label{eq:eta-dynamic-higgs}
\end{equation}
where $g_{\rm eff}$ is an effective coupling which encodes how efficiently the hot plasma transfers momentum to the wall and $T_{\rm eff}(\tau)$ is an
effective temperature combining a small ambient seed temperature with the
dynamically generated shock temperature in the thermal shell:
\begin{equation}
T_{\rm eff}^4(\tau)
=
T_{\rm amb}^4
+ T_{\rm shock}^4(\tau),
\qquad
T_{\rm shock}^4(\tau)
=
\frac{30}{\pi^2 g_*}\,\rho_{\rm th}(\tau).
\end{equation}
This form is suggested by simple kinetic theory: the drag is proportional to
the momentum flux carried by thermal particles, which scales as
$n\,p \sim T^3 \times T \sim T^4$, and to the probability $\sim g_{\rm eff}^2$
that these particles scatter and transfer their momentum to the wall. 

The frictional pressure entering the wall equation of motion is $P_{\rm fric}(\tau)\equiv \eta(\tau)\,\gamma(\tau)v(\tau);$ with the temperature–dependent ansatz $\eta(\tau)=g_{\rm eff}^2\,T_{\rm eff}^4(\tau)$ this becomes
\begin{equation}
P_{\rm fric}(\tau)=g_{\rm eff}^2\,\gamma(\tau)v(\tau)\,T_{\rm eff}^4(\tau),
\end{equation}
consistent with ultrarelativistic bubble–wall friction calculations in which the thermal backreaction is controlled by the local plasma temperature together with the wall boost (up to coupling and phase–space factors) \cite{Long:2024sqg,Hoche:2020ysm,BarrosoMancha:2020fay}.  Parametrically,
this is consistent with detailed calculations of electroweak bubble friction,
where the thermal pressure (and thus the effective drag) on an ultra--relativistic
wall scales as $P_{\rm th}\propto \gamma^{n}\,T^4$ (up to coupling/phase-space factors), with $n$ depending on the dominant kinematics and emission spectrum, see e.g. the all–orders analysis of Ref.\cite{Long:2024sqg, Hoche:2020ysm, BarrosoMancha:2020fay}.

\paragraph{Relativistic scaling of the drag.}
\label{Scaling_of_the_drag}
Following the ultrarelativistic discussion of Long \& Turner\cite{Long:2024sqg}, the thermal pressure
on the wall may be schematically written in the wall frame as
\(P_{\rm th}\sim F_a\langle\Delta p_z\rangle\), where the incident flux scales as
\(F_a\sim \gamma \,T^3\).
For the classic \(1\!\to\!1\) kinematics (mass change across the wall with no emission),
one has \(\Delta p_z\propto 1/E_a\sim 1/(\gamma \,T)\), so \(P_{\rm th}\propto \gamma^0\).
Including soft/collinear \(1\!\to\!2\) emission changes the momentum transfer: for massive
emission channels the transfer can be set by the emitted mass scale, yielding
\(P_{\rm th}\propto \gamma^1\).
By contrast, if the emission spectrum is approximately log-flat up to a UV cutoff of order
the incident energy \(E_a\sim \gamma\, T\), then \(\langle\Delta p_z\rangle\sim E_a\) and the
pressure can scale as \(P_{\rm th}\propto \gamma^2\).

Motivated by this structure, we adopt the phenomenological drag
which explicitly realizes the \(\gamma^1\) scaling associated with soft/collinear emission
in massive channels, while packaging microphysical coupling/phase-space factors into
\(g_{\rm eff}^2\) and the local energy density into \(T_{\rm eff}^4\).
Contributions that can enhance the scaling toward \(\gamma^2\)---e.g. harder (UV-dominated)
emission and approximately log-flat spectra extending up to a UV cutoff---are not included
in our baseline model. Our quoted thermal yields should therefore be interpreted as
conservative in the ultrarelativistic regime.

To connect the dynamic ansatz \eqref{eq:eta-dynamic-higgs} to the terminal
condition \eqref{eq:eta-target-higgs}, we fix, for each $\delta$, the ambient
temperature $T_{\rm amb}$ by demanding that at nucleation
($\rho_{\rm th}=0$, $T_{\rm shock}=0$) the instantaneous drag reproduces
$\eta(0)$:
\begin{equation}
\eta(0)
=
g_{\rm eff}^2\,T_{\rm eff}^4(0)
=
g_{\rm eff}^2\,T_{\rm amb}^4,
\end{equation}
which yields
\begin{equation}
T_{\rm amb}
=
\left[\frac{\eta(0)}{g_{\rm eff}^2}\right]^{1/4}.
\end{equation}
In our numerical Higgs benchmarks we adopt
\begin{equation}
g_{\rm eff} = 10^{-3},
\end{equation}
So that the friction starts from the constant–$\eta$ value implied by the chosen $\delta$ and then grows self-consistently as the shocked layer heats up. In this way, the drag self-regulates: as $T_{\rm eff}$ rises the friction increases, providing negative feedback that curbs further heating, while remaining anchored to the terminal–velocity deficit set by $\delta$.

The thermal energy density and multiplicity are updated using the energy--deficit law\footnote{Note that while the friction pressure scales as $P_{\rm fric}=\eta\,\gamma v,$ the energy deposition rate entering $d\rho_{\rm th}/d\tau$ scales as work done per unit time, hence carries an additional factor $\sim \gamma v,$ giving a contribution $\propto \eta\,\gamma^2 v^2.$}
\begin{equation}
\frac{d\rho_{\rm th}}{d\tau}
=
\gamma\,\mu\,\big(v_0 - v\big)
\left(\Delta V - \frac{2\sigma}{R}\right)
+ \eta(\tau)\,\mu
\gamma^2 v^2,
\label{eq:rho-evolution-final}
\end{equation}
combined with the usual equilibrium relations
\begin{equation}
T(\tau)
=
\left(\frac{30}{\pi^2 g_*}\,\rho_{\rm th}(\tau)\right)^{1/4},
\qquad
n_{\rm th}(\tau)
=
\frac{\zeta(3)}{\pi^2}\,g_*\,T^3(\tau),
\end{equation}
and the comoving production rate
\begin{equation}
\frac{dN_{\rm th}}{d\tau}
=
4\pi R^2(\tau)\,\sinh y(\tau)\;n_{\rm th}(\tau).
\end{equation}
Here $\mu$ is the Higgs--like mass scale in the true vacuum and $g_*=106.75$.

\vspace{2mm}
\noindent\textbf{Integration time.}
It is convenient to introduce the constant--friction reference timescale
\begin{equation}
\tau_{\rm term}
=
\frac{\sigma}{\eta(0)},
\end{equation}
which is the natural proper time controlling the approach to terminal velocity
in the auxiliary system with fixed drag $\eta(0)$.  In
that simplified model the solution of the linearised equation of motion from Eqn~\eqref{eq:friction-eom} in the late time scenario for a large bubble (neglecting the curvature term) shows
an approximately exponential approach to the asymptotic value,
\begin{equation}
\gamma(\tau)v(\tau)
\simeq
\gamma_{\rm term}v_{\rm term}
\Bigl(1-e^{-\tau/\tau_{\rm term}}\Bigr),
\end{equation}
so that by time~$\tau$ a fraction $1-e^{-\tau/\tau_{\rm term}}$ of the asymptotic
impulse has been accumulated.

In our full simulations the drag coefficient is not constant but evolves
according to Eq.~\eqref{eq:eta-dynamic-higgs}.
The quantity $\tau_{\rm term}$ should therefore be interpreted as a convenient
normalisation inherited from the constant–$\eta$ limit, rather than as an exact
fit to the fully non–linear dynamics.  We choose the total integration time as
\begin{equation}
\tau_{\rm final}
=k\,\tau_{\rm term},
\end{equation}
with
\begin{equation}
k = 5,
\qquad
\tau_{\rm final} = 5\,\tau_{\rm term}.
\end{equation}
In the constant–friction toy model this would capture more than $99\%$ of the
asymptotic energy deficit, since $1-e^{-5}\simeq 0.993$, while in the full
dynamic–$\eta$ evolution the increasing friction causes the wall to reach its
quasi–terminal regime even faster.  We have checked explicitly that increasing
$k$ from $5$ to $6$ changes the final thermal multiplicity
$N_{\rm th}$ only at the few--percent level, whereas choosing
$k=3$ underestimates it at the tens of percent level.  The
choice $\tau_{\rm final}=5\,\tau_{\rm term}$ is therefore a conservative and
numerically stable benchmark to quote Higgs--channel thermal yields. A detailed justification of the integration cutoffs, together with quantitative convergence checks, is provided in Appendix~\ref{App-C}.

Finally, we do not extend the integration arbitrarily far beyond
$\tau_{\rm final}$, even though in principle an expanding bubble could
continue to radiate and produce particles as long as it cruises near its
terminal speed.  Our estimate is explicitly local and comoving: the shell is
moving relativistically, the radiation it emits is boosted and streams out
until it leaves the local horizon, and this outgoing flux induces a
radiation–reaction (self–force) on the wall.  At the same time, ambient
particles scatter off the shell, absorb energy, and are kicked away.  These
dissipative channels provide additional sources of friction, effectively
encoded in the last term of Eq.~\eqref{eq:rho-evolution-final}, and are not
contained in the surface tension alone.  In practice we find that evolving up
to $\tau_{\rm final}=5\,\tau_{\rm term}$ already captures the dominant part of
the thermal production: extending the evolution further changes $N_{\rm th}$
only mildly within our present idealised setup, while truly reliable late--time
behaviour would require including additional physics --- strong gravity,
hydrodynamic backreaction of the shocked plasma, and detailed microphysics of
radiation reaction and scattering.  Our finite $\tau_{\rm final}$ should
therefore be viewed as a conservative upper bound on the thermal multiplicity
within the regime where our approximations remain under quantitative control.
\subsection*{Numerical results}

For the Higgs false--vacuum benchmark we take
\[
\Delta V = 5.869\times10^{-12} M_p^4,
\qquad
R_0 = 2\times10^{4} M_p^{-1},
\qquad
\sigma = 5.8680\times10^{-8} M_p^3,
\]
with $g_*=106.75$ and $\zeta(3)=1.202056$.  The true--vacuum Higgs--like mass
scale is
\begin{equation}
\mu = 7.16\times10^{-4} M_p,
\end{equation}
which sets the shocked--layer thickness $\ell\sim 1/\mu$ in the heating term.
We consider three ultra--relativistic terminal--velocity deficits,
\[
\delta = 10^{-8},\qquad 10^{-9},\qquad 10^{-10},
\]
compute $\eta(0)$ from Eq.~\eqref{eq:eta-target-higgs}, fix
$T_{\rm amb}(\delta)$ as above, and evolve the coupled system with the dynamic
drag coefficient \eqref{eq:eta-dynamic-higgs} up to
\begin{equation}
\tau_{\rm final}=5\,\tau_{\rm term},\qquad 
\tau_{\rm term}\equiv \frac{\sigma}{\eta(0)}.
\end{equation}
For each run we integrate the proper--time equations
$dR/d\tau=\sinh y$ and $dy/d\tau=A-2/R-(\eta/\sigma)\sinh y$ with
$A\equiv \Delta V/\sigma$, and compute the thermal number from
$dN_{\rm th}/d\tau=(dV/d\tau)\,n(T)$ with $dV/d\tau=4\pi R^2\sinh y$.
The resulting thermal yields are summarised in Table~\ref{tab:thermal-yields}.

\begin{table}[h]
\centering
\small
\begin{tabular}{@{}lccc@{}}
\toprule
Scenario ($\delta$) 
& $\eta(0)\,[M_p^4]$ 
& $\tau_{\rm final}\,[M_p^{-1}]$
& $N_{\rm th}$ \\
\midrule
$10^{-8}$  &
$8.30\times10^{-16}$ &
$3.53\times10^{8}$ &
$2.15\times10^{26}$ \\
$10^{-9}$  &
$2.62\times10^{-16}$ &
$1.12\times10^{9}$ &
$1.87\times10^{27}$ \\
$10^{-10}$ &
$8.30\times10^{-17}$ &
$3.53\times10^{9}$ &
$1.61\times10^{28}$ \\
\bottomrule
\end{tabular}
\caption{
Thermal Higgs--channel particle yields obtained from the energy--deficit
formulation with a temperature--dependent drag coefficient
$\eta(\tau)=g_{\rm eff}^2 T_{\rm eff}^4(\tau)$ and $g_{\rm eff}=10^{-3}$.
For each $\delta$, the reference value $\eta(0)$ is fixed
by the terminal condition $\Delta V=\eta(0)\gamma_{\rm term}v_{\rm term}$,
and the evolution is integrated up to $\tau_{\rm final}=5\,\tau_{\rm term}$.}
\label{tab:thermal-yields}
\end{table}

For fixed $\Delta V$, the terminal matching implies $\eta(0)\propto
(\gamma_{\rm term}v_{\rm term})^{-1}$, so $\eta(0)$ decreases as $\delta$
is reduced.  In the fully dynamical system this does \emph{not} imply a smaller
thermal yield: weaker friction allows the wall to expand over a larger
four--volume before heating shuts off, and the integrated swept volume
$\int d\tau\,4\pi R^2\sinh y$ grows strongly with decreasing $\delta$.
In our benchmarks the thermally produced multiplicities are
\begin{equation}
N_{\rm th} \sim 10^{26}\text{--}10^{28},
\end{equation}
which are enormous on particle--physics scales.

If we interpret these quanta as true--vacuum Higgs excitations of mass
$m_H \simeq \mu = 7.16\times 10^{-4} M_p$, the total decay energy stored in the
hot shell is
\begin{equation}
E_{\rm tot} \sim N_{\rm th}\, m_H .
\end{equation}
If a fraction $f_\nu$ of this energy is ultimately channelled into
$\mathcal{O}(1$--$100)\,\mathrm{GeV}$ neutrinos (and similarly into photons),
the corresponding neutrino multiplicity is roughly
\begin{equation}
N_\nu 
\sim 
f_\nu\,N_{\rm th}\,
\frac{m_H}{\langle E_\nu\rangle}
\sim
10^{37}\text{--}10^{39}\,
\left(\frac{f_\nu}{10^{-2}}\right)
\left(\frac{10~\mathrm{GeV}}{\langle E_\nu\rangle}\right),
\end{equation}
for the range of $N_{\rm th}$ in Table~\ref{tab:thermal-yields}.  Although
this is many orders of magnitude below the neutrino yield of a core--collapse
supernova at the source, the absolute number of quanta is still macroscopically
large.  For a bubble nucleated within $\mathcal{O}(10)\,\mathrm{pc}$ of the
Earth such multiplicities correspond to $\mathcal{O}(10^3$--$10^7)$ neutrinos
crossing a km$^2$--scale detector, so a nearby late--time phase--transition
event would appear as an exceptionally bright, short--duration burst in
high--energy neutrinos and photons.  At cosmological distances, of order
$10^6$–$10^8$ light years, the same total energy released in very
high--energy photons would manifest as an extremely luminous, pointlike
transient --- a ``shiny'' high--energy source on the sky.
\subsection{Thermal spectrum of the produced particles}

Once the total thermal energy and particle number have been determined
from the dynamical evolution, the phase–space distribution of the produced
particles is fixed by equilibrium thermodynamics.  
Since the thermal bath is generated in a thin shocked layer and rapidly equilibrates,
the momentum distribution of any light bosonic species is well described by the
massive Bose–Einstein spectrum
\begin{equation}
f(k) \equiv \frac{k^{2}}{\exp\!\left(\sqrt{k^{2}+m^{2}}/T\right)-1},
\label{eq:massiveBE}
\end{equation}
where $m$ is the mass of the scalar produced (for the Higgs-like field used in
our benchmark, $\mu = 7.16\times 10^{-4} M_p$) and $T$ is the thermalisation
temperature extracted from the shocked plasma.

\begin{figure}[h]
    \centering
    \includegraphics[width=0.75\textwidth]{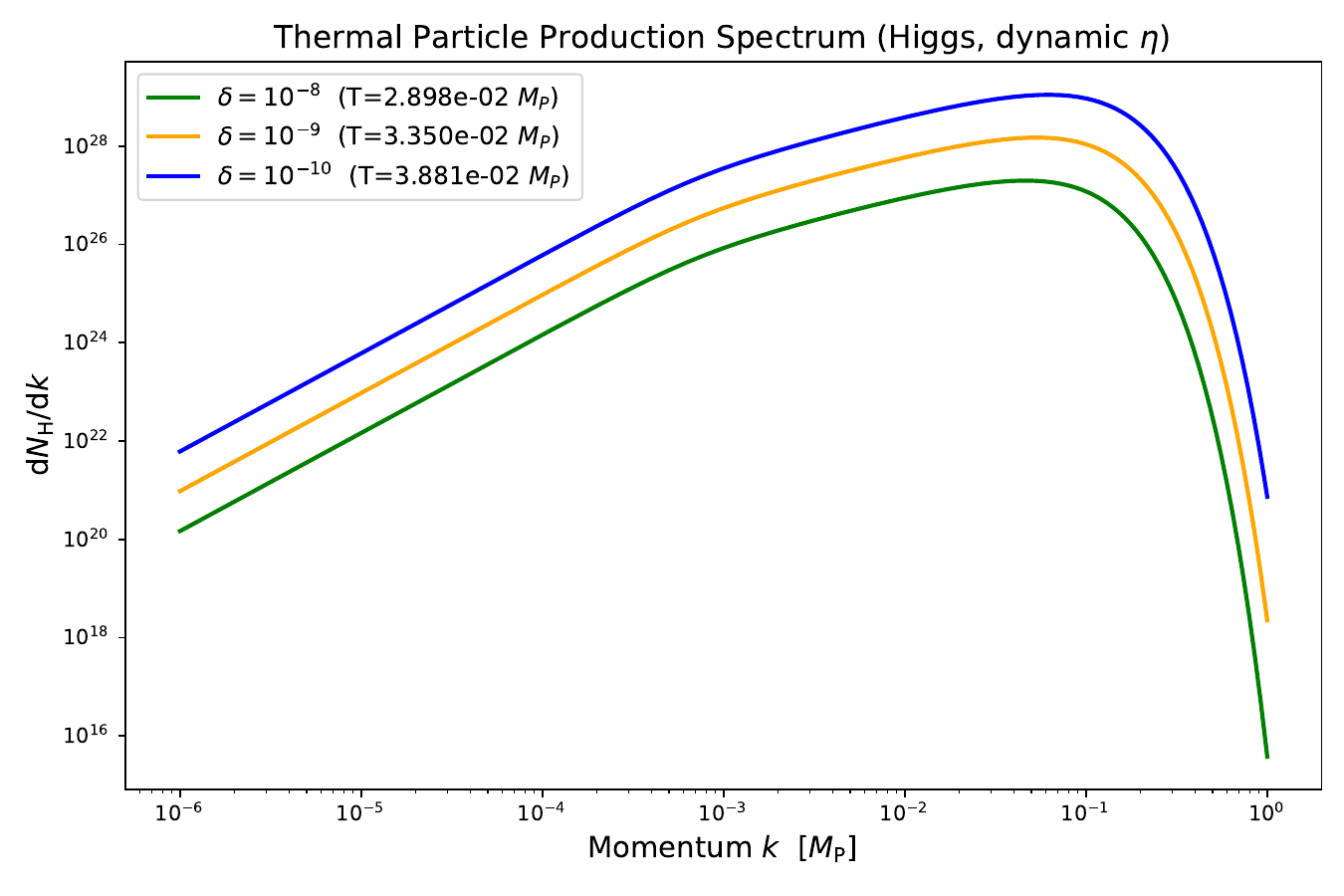}
    \caption{
        Physically normalised thermal spectrum ${\rm d}N/{\rm d}k$ for the
        Higgs-like scalar with mass 
        $\mu = 7.16\times10^{-4}M_p$ produced via frictional dissipation.  
        Each curve corresponds to a different terminal-velocity deficit
        $\delta$, which determines both the shock temperature and the total
        thermal particle number $N_{\rm th}$.  
        The peak of each distribution occurs at $k_{\rm peak}\simeq 2.8T$, as
        expected for a massive Bose–Einstein spectrum.  
        The vertical normalisation of the curves reflects the physical
        particle yields shown in Table~\ref{tab:thermal-yields}.  
    }
    \label{fig:thermal-spectrum}
\end{figure}

\vspace{2mm}
\noindent\textbf{Normalisation to the physical yield.}
The shape of the spectrum is fixed by (\ref{eq:massiveBE}),  
but the overall normalisation must reproduce the total thermal particle number
$N_{\rm th}$ obtained from the integrated heating history in
Eq.~\eqref{eq:dNth_final_v2}.  
We therefore define the \emph{physically normalised} spectrum
\begin{equation}
\frac{{\rm d}N}{{\rm d}k}
=
N_{\rm th}\;
\frac{f(k)}{\displaystyle \int_0^\infty f(k)\,{\rm d}k},
\label{eq:normalised-spectrum}
\end{equation}
which guarantees
\begin{equation}
\int_0^\infty \frac{{\rm d}N}{{\rm d}k}\,{\rm d}k = N_{\rm th}.
\end{equation}
This ensures that the curves shown in Fig.~\ref{fig:thermal-spectrum}  
represent the correct physical particle numbers generated by frictional
dissipation for each value of the terminal-velocity deficit~$\delta$.

\vspace{2mm}
\noindent\textbf{Peak structure and scaling.}
For bosonic thermal distributions with $m \ll T$, the maximum of
(\ref{eq:massiveBE}) occurs very close to
\begin{equation}
k_{\rm peak} \simeq 2.8\,T,
\label{eq:kpeak-scaling}
\end{equation}
a standard result that remains accurate for the Higgs-like mass used here.
Because smaller~$\delta$ leads to much larger thermal energy densities and thus
larger temperatures $T$, the peak of the spectrum shifts systematically to
higher momenta as $\delta$ decreases.  
This scaling, together with the large hierarchy in total particle numbers,
explains the structure and vertical separation of the three curves in the
final spectrum.

\vspace{2mm}
\noindent\textbf{Resulting thermal spectrum.}
Figure~\ref{fig:thermal-spectrum} displays the physically correct, fully
normalised spectrum ${\rm d}N/{\rm d}k$ for the three choices of~$\delta$
considered in this work.  
Each curve encodes:
(i) the massive Bose–Einstein shape determined by the shock temperature,
(ii) the shift of the peak at $k_{\rm peak}\simeq 2.8T$, and 
(iii) the correct total particle yield fixed by~$N_{\rm th}$.
The steep decline at $k\gg T$ reflects Boltzmann suppression in the massive tail
of the distribution.  
Although the plot explicitly shows the spectrum for the Higgs scalar,
the same thermal bath also populates all lighter Standard Model degrees of freedom
with comparable characteristic momenta $k\sim T$.

These results demonstrate that thermalization of the dissipated energy dominates overwhelmingly over vacuum-mismatch particle emission when the terminal-velocity deficit is small.  
A tiny friction coefficient allows the wall to reach extreme Lorentz factors and sustain a long relaxation period, during which the deficit between the physical and frictionless trajectories is gradually converted into heat.  
The energy-deficit formulation ensures that this conversion is computed correctly and because the energy-deficit approach tracks the exact portion of wall energy that must be dissipated irrespective of microscopic uncertainties, our analysis yields a robust and model-independent estimate of thermal particle production.

The enormous population of Higgs particles produced through this thermal channel is not stable: each Higgs boson rapidly decays into Standard Model final states, with dominant branching fractions into photons, neutrinos, and charged leptons. Once emitted, the photons free–stream across interstellar and intergalactic distances essentially unimpeded, while the accompanying neutrinos traverse the cosmos with negligible attenuation. Consequently, the Higgs decays generate a distinctive observational signature at Earth in the form of transient gamma-ray bursts and correlated high-energy neutrino bursts. Existing and upcoming gamma-ray detectors and future radio-Cherenkov neutrino arrays—operate in the energy and time windows relevant for short-duration, broadband photon and neutrino bursts of this type.  However, because these
messengers dilute as $1/d^2$ and must compete with astrophysical backgrounds, an observable signal is expected only for bubbles nucleated at relatively short distances on astrophysical scales. By contrast, the violent acceleration and shock heating of the wall also source
gravitational radiation whose strain decays more slowly and can remain detectable over cosmological distances.  For suitable choices of parameters, the resulting high-frequency burst could fall within the sensitivity reach of space-based interferometers (e.g.~LISA) and future facilities such as Cosmic Explorer and the Einstein Telescope.  Thus, the thermal production mechanism does not merely represent an internal energy-loss channel of the bubble wall: it implies a concrete, though likely rare, multi-messenger target, in which photons and neutrinos probe sufficiently nearby bubbles while the associated gravitational-wave burst can, in principle, be seen from much farther away, with the overall signal strength directly tied to the ultra-relativistic dynamics of the transition.

\subsection{Sub-Planckian temperature regime.}
A related conceptual issue is the very high shock temperatures that the model could in principle generate.  
Even though $\Delta V$ is fixed and local energy conservation is enforced, the total vacuum energy
available to the wall is not bounded: as the bubble expands it converts an ever larger volume of
false vacuum, so the released energy grows as $\sim \Delta V\,R^3$.  
If all of this work were deposited into an infinitesimally thin shell of fixed thickness, the local
thermal energy density would grow roughly as $\rho_{\rm th}\propto \Delta V\,R$ and could formally
approach or exceed the Planck scale.

In practice, once $\rho_{\rm th}$ approaches the fundamental high scale of the problem (and in
particular $M_p^4$) our effective field-theory and flat-space descriptions cease to be reliable:
strong-gravity effects, gravitational backreaction, and possibly new microscopic degrees of freedom
are expected to become important, and one may even encounter scenarios of shell-like gravitational
collapse or black–membrane–like configurations\cite{Kibble:1976sj,Kibble:1980mv,Hoyos:2026eoq,Bea:2022mfb, Vachaspati:2006zz, Rajantie:2003xh, Enqvist:1991xw, Gergely:2006ei, Sadeghi:2015vaa,Wong:2010rg,Harko:2004ui,Bruni:2001fd,Horowitz:2001cz,Gregory:1993vy,Gregory:2000gf,Chamblin:1999by,Garriga:1999yh,Majumdar:2005ba,Dadhich:2000am}.  A faithful treatment of such a regime would require
including dynamical gravity and going beyond the simple hydrodynamic + thin-wall approximation
employed here.\footnote{One could refer to these articles for a detailed hydrodynamic treatment relevant to our study. Ref~\cite{Ai:2024shx,Ai:2025bjw,Konstandin:2010cd,Konstandin:2010dm,Bea:2024bxu,Krajewski:2023clt,Krajewski:2024gma,Wang:2022txy,Wang:2023kux,DeCurtis:2022hlx,DeCurtis:2024hvh, Attems:2017ezz}}

In the present work we deliberately stay away from this strongly coupled, Planckian regime.
With the choice $\ell\simeq1/\mu$ (corresponding to a comoving layer of thickness
$\ell\sim 1/\mu$) and with our benchmark values of $\Delta V$, $\sigma$ and $\delta$, the
numerical evolution up to $\tau_{\rm final}=5\,\tau_{\rm term}$ always yields
sub-Planckian energy densities and temperatures, $\,\rho_{\rm th}\ll M_p^4$ and
$T(\tau_{\rm final})\ll M_p$.  We therefore do \emph{not} need to impose an explicit hard cutoff in
$T$ in the runs shown in Fig.~\ref{fig:thermal-spectrum}: all quoted thermal yields are obtained
within a regime where the EFT and flat-space approximations remain under quantitative control.

If one were to explore more extreme regions of parameter space in which the formal evolution drives
$T(\tau)$ close to $M_p$, the correct description would have to be modified at that point, either by
introducing an explicit sub-Planckian cutoff or, more ambitiously, by coupling the wall–plasma
system to full gravitational dynamics and possible new high-scale physics.  Such Planckian-shell
scenarios are intriguing in their own right, but lie beyond the scope of the present analysis and
are left for future work.

Taken together, the choices of $\tau_{\rm final}$ and of the comoving layer prescription
$\ell\sim 1/\mu$ ensure that our quoted thermal yields constitute a conservative bound
computed entirely in a sub-Planckian regime, while avoiding spurious contributions from late-time or
super-Planckian dynamics that our simplified framework is not designed to describe.

\section{Signal lead time}
\label{sec:signal_leadtime}

If expanding bubble walls propagate with a terminal velocity slightly below the speed of light, secondary particles such as photons or neutrinos emitted from Higgs decays will reach distant observers in advance of the wall itself. In this sense, they may act as an early ``doomsday'' warning signal. We now quantify the expected delay.

\subsection{Cosmological inputs}
We adopt a flat $\Lambda$CDM background with present-day parameters
\begin{align}
H_0 &= \frac{67.4\times10^3\;\mathrm{m\,s^{-1}}}{3.0857\times10^{22}\;\mathrm{m}}
\;\approx\;2.184\times10^{-18}\;\mathrm{s^{-1}}, \\[4pt]
c   &= 2.99792458\times10^8\;\mathrm{m\,s^{-1}},\qquad
\Omega_m=0.315,\qquad
\Omega_\Lambda=0.685.
\end{align}
The Hubble rate evolves with redshift as
\begin{equation}
H(z)=H_0\sqrt{\Omega_m(1+z)^3+\Omega_\Lambda}.
\label{eq:Hz}
\end{equation}
The corresponding comoving distance is
\begin{equation}
\chi(z)=\int_{0}^{z}\frac{c}{H(z')}\,dz',
\label{eq:chi}
\end{equation}
while the photon travel time as measured by a comoving observer is
\begin{equation}
t_\gamma(z)=\int_{0}^{z}\frac{dz'}{(1+z')\,H(z')}.
\label{eq:tgamma}
\end{equation}
For illustration, we choose a source at proper distance 
\begin{equation}
D_{\rm ly}=1.0\times10^8\;\mathrm{ly}, 
\quad
D_m=D_{\rm ly}\times9.4607\times10^{15}\;\mathrm{m},
\end{equation}
and determine the corresponding emission redshift by solving
\begin{equation}
\chi\!\left(z_{\rm em}\right)=D_m,
\label{eq:zsolve}
\end{equation}
which gives $z_{\rm em}\simeq0.0069$.

\subsection{Arrival delay for bubble walls}
Consider now a bubble expanding at velocity
\begin{equation}
v=(1-\delta)c, \qquad \delta \ll 1.
\end{equation}
In comoving coordinates the wall trajectory satisfies $d\chi/d\eta = 1-\delta$, so the arrival delay relative to photons can be expressed in full generality as
\begin{equation}
\Delta t(z) = \frac{\delta}{1-\delta}\int_{0}^{z}\frac{dz'}{H(z')}.
\label{eq:delay_general}
\end{equation}
This expression is valid at arbitrary redshift, and naturally reduces to the Minkowski estimate at low $z$. Expanding $H(z)\simeq H_0$ for $z\ll 1$ yields
\begin{equation}
\Delta t \;\simeq\; \delta\,\frac{D}{c},
\label{eq:delay_flat}
\end{equation}
with $D$ the proper distance today. This approximation suffices for the illustrative case $D=10^8$\,ly ($z\simeq0.007$).

\begin{table}[h!]
\centering
\begin{tabular}{|ccc|}
\hline
\textbf{Velocity Deficit $\delta$} & \textbf{Distance (ly)} & \textbf{Time Delay} \\
\hline
$1.0\times10^{-8}$  & $1.0\times10^8$ & 1 yr, 0 d, 5 h, 58 m, 18.35 s \\
$1.0\times10^{-9}$  & $1.0\times10^8$ & 0 yr, 36 d, 12 h, 35 m, 49.84 s \\
$1.0\times10^{-10}$ & $1.0\times10^8$ & 0 yr, 3 d, 15 h, 39 m, 34.98 s \\
\hline
\end{tabular}
\caption{Arrival delays for bubble walls with subluminal deficits $\delta$ relative to photons over a distance of $1.0\times10^8$ light years, using the flat-space approximation $\Delta t \simeq \delta D/c$. For higher redshifts, Eq.~\eqref{eq:delay_general} must be used.}
\label{tab:photon_delay}
\end{table}

Even a minute shortfall from luminal expansion produces a tangible lead time: photons or neutrinos emitted in Higgs decays would arrive days to months ahead of the wall itself. Such precursors could in principle serve as advance warnings of a cosmological true vacuum decay event. The three cases shown in Table~\ref{tab:photon_delay} are purely illustrative; other choices of $\delta$ and redshift are straightforwardly evaluated using Eq.~\eqref{eq:delay_general}.

\section{Conclusions}
The work presented here was motivated by an interplay between the LHC collider data and recent theoretical progress in the field theory in curved spacetime. The standard model precision data coming from the LHC indicate that we currently live in a false vacuum. While the LHC data indicate that the lifetime of the false vacuum is safely large, it has been shown that small primordial black holes can serve as the phase transition catalysers  and significantly increase the transition probability. In fact, it might be possible that the bubbles of the true vacuum already exist in our visible universe. One of the signatures of the existence of such bubbles would be copious production of the heavy Higgs particles due to the vacuum mismatch inside and outside of the bubble. We used the standard field theory techniques to calculate the spectrum of Higgs particles produced this way. Ideally, one would want to use the event generators to simulate the Higgs decay and subsequent evolution of the decay products. However, with the current state-of-the-art in the field, this is not possible. The mass of the Higgs particle in the new vacuum is very large ($\sim 10^{15}\,\mathrm{GeV}).$ At these scales, the perturbation theory certainly breaks down, and the results from the standard event generators cannot be extrapolated. Moreover, there are indications that far above the electroweak breaking scale the Higgs does not even behave as a well defined particle, but rather as a resonance \cite{Dai:2014xka}. Fortunately, we can reasonably argue that most of the energy invested into the Higgs particle production will eventually end up in photons and neutrinos. The reason is that the Higgs is a neutral particle and will decay into an equal number of particles and antiparticles. Some of them will be unstable and will further decay, and some of them will hadronize and then decay. As the final product we will get mostly photons, neutrinos, leptons and anti-leptons. Some of these leptons and anti-leptons will annihilate giving again photons. 
Thus, most of the energy will go into a long-range signature - photons and neutrinos. 

The bubble wall's velocity as seen by an external observer typically evolves in time as $v_b \sim t/\sqrt{t^2+R_0^2} $. Thus, for $t \gg R_0$, this velocity is almost the speed of light. 
If this remains true at all times, a very large number of particles will be produced due to the vacuum mismatch since the proper acceleration never ceases, but we will have no chance to observe them before the wall hits us.  However, the wall velocity is never exactly equal to the speed of light. In fact, detailed calculations indicate that the bubble wall velocity might be limited to well below the speed of light  \cite{Branchina:2025jou}.
In addition, the bubble wall can be viewed as a highly coherent state of a large number of the Higgs quanta. As such, it will strongly interact with virtually all the standard model particles. In the early universe, it will interact with the surrounding plasma, while at late times it will interact with other bubbles, stars, planets, interstellar and intergalactic gas, and finally even with the particles it creates itself. These interactions would further slow down the propagation of the wall. It is thus not very unreasonable to expect that these photons and neutrinos should be able to reach us before the bubble wall. 

If the friction with the environment exists, then the proper acceleration will cease at some point. Since the particle production due to vacuum-mismatch mechanism is analogous to an Unruh-type emission that operates only while the bubble wall is accelerating, our calculated  spectrum accounts for particle production only up to the point where the wall attains its terminal velocity. After acceleration ceases, the mismatch channel shuts off. Nevertheless, the wall continues to transfer enormous energy into the environment through friction during the terminal phase, and this energy is rapidly thermalized in the shocked layer behind the wall. As we have shown, this thermalization process itself becomes an efficient source of additional particles, complementing the vacuum-mismatch production even after the wall has stopped accelerating. This process produces an enormous amount of particles in a very short period of time, which means that the period of friction does not have to last for very long. For our conservative benchmarks the total neutrino yield is some orders of magnitude smaller than that of a typical core–collapse supernova, but the characteristic energies are much higher, and a bubble nucleated very close to us would still appear as an exceptionally bright, short–duration burst. 

If friction continues for some extended period of time, a huge vacuum energy density which is converted into thermal radiation may exceed the Planckian energy density in a thin layer around the bubble. In that case, the resulting spacetime geometry may likely resemble a (spherical) black membrane. However, such extended objects are unstable and will most likely decay into a large number of smaller black holes. If that indeed happens, then the phase transition might effectively be quenched, and the signals might reach us without the subsequent catastrophe.

In a broader context, a late time first‐order electroweak transition is expected to leave a multi‐messenger imprint— possibly a stochastic gravitational‐wave background, subtle distortions in the CMB and 21-cm line, and potential signals in gamma-ray, cosmic-ray and pulsar-timing data—that upcoming facilities are well poised to detect.  These direct and indirect channels provide the most robust avenues for detecting the bubbles of the true vacuum.  In a follow-up study, we will explore the detectability and present detailed forecasts for both current and next-generation observatories.

We note that the crucial assumption here is that there is no new physics beyond the standard model. New physics could potentially modify the Higgs potential and render our vacuum stable. Another possibility is that small primordial black holes do not exist. Since they are necessary catalysers, their absence in an appropriate mass range \cite{Dai:2019eei} would perhaps invalidate this doomsday scenario. We also note that our quantitative benchmark depends on the location of the stabilized true minimum of the Higgs potential.  If the true-vacuum scale were lower than in the benchmark adopted here, then all characteristic mass scales in the problem would be correspondingly reduced.  In particular, the Higgs mass in the true vacuum would be smaller, the bubble wall thickness would typically be larger, and the resulting photon and neutrino spectra would be shifted toward lower characteristic energies.  The qualitative picture developed in this work would remain unchanged --- namely, bubble expansion, vacuum-mismatch particle production, friction-induced thermal production, and long-range photon/neutrino signatures --- but the signal would be softer and the detailed phenomenology would have to be re-evaluated for that lower-scale vacuum structure. We also stress that even if Higgs-instability regions are seeded in the Universe, their fate need not be “doomsday” in all circumstances: De Luca--Kehagias--Riotto \cite{DeLuca:2022cus} argue that in the thin-wall regime such dangerous patches can collapse due to Higgs backreaction (often leaving black holes), rather than expanding and eating the Universe, which can substantially relax the corresponding inflationary Hubble bound.

\begin{acknowledgments}
The authors are grateful to D. Wackeroth and C. Williams for their help with event generators and very useful comments on the paper. D.C. Dai is supported by the National Science and
Technology Council (under grant no. 111-2112-M-259-
016-MY3). AS and DS are partially supported by the US National
Science Foundation, under the Grant No. PHY-2310363.
\end{acknowledgments}
\appendix
\section{Appendix~A: Proper Acceleration of the Bubble Wall}\label{App-A}

We derive here the magnitude of the proper acceleration of the bubble wall described by the equation \( r^2 - t^2 = R_0^2 \).

The bubble wall follows a hyperbolic trajectory in \((t, r)\) spacetime, which can be parametrized using proper time \(\tau\) (the time experienced by an observer on the bubble wall). Note that this $\tau$ is not the same as the Euclidean time  $\tilde{\tau}$ in Section \ref{pp}. The proper parametrization is:
\begin{equation}
t = R_0 \sinh\left(\frac{\tau}{R_0}\right), \quad r = R_0 \cosh\left(\frac{\tau}{R_0}\right).
\end{equation}
This satisfies the equation \( r^2 - t^2 = R_0^2 \), as:
\begin{equation}
R_0^2 \cosh^2\left(\frac{\tau}{R_0}\right) - R_0^2 \sinh^2\left(\frac{\tau}{R_0}\right) = R_0^2.
\end{equation}

The 4-velocity \( u^\mu \) of the bubble wall is the derivative of the coordinates with respect to proper time
\begin{equation}
u^\mu = \left( \frac{dt}{d\tau}, \frac{dr}{d\tau} \right) = \left( \cosh\left(\frac{\tau}{R_0}\right), \sinh\left(\frac{\tau}{R_0}\right) \right).
\end{equation}
Thus, the norm of the 4-velocity is
\begin{equation}
u^\mu u_\mu = \left(\frac{dt}{d\tau}\right)^2 - \left(\frac{dr}{d\tau}\right)^2 = \cosh^2\left(\frac{\tau}{R_0}\right) - \sinh^2\left(\frac{\tau}{R_0}\right) = 1,
\end{equation}
confirming that \(\tau\) is the proper time.

The 4-acceleration \( a^\mu \) is the derivative of the 4-velocity with respect to proper time
\begin{equation}
a^\mu = \frac{du^\mu}{d\tau} = \left( \frac{1}{R_0} \sinh\left(\frac{\tau}{R_0}\right), \frac{1}{R_0} \cosh\left(\frac{\tau}{R_0}\right) \right).
\end{equation}
The norm of the 4-acceleration (which gives the proper acceleration) is
\begin{equation}
a^\mu a_\mu = \left(\frac{1}{R_0} \sinh\left(\frac{\tau}{R_0}\right)\right)^2 - \left(\frac{1}{R_0} \cosh\left(\frac{\tau}{R_0}\right)\right)^2 = -\frac{1}{R_0^2}.
\end{equation}
The magnitude of the proper acceleration is the absolute value of the norm
\begin{equation}
|a| = \sqrt{|a^\mu a_\mu|} = \frac{1}{R_0}.
\end{equation}

Thus, in the absence of friction, the magnitude of the proper acceleration of the bubble wall is constant and given by \cite{Good:2013lca,Walker:1984vj}
\begin{equation}
\dfrac{1}{R_0}.
\end{equation}

Since the relationship between proper and coordinate acceleration is $a =\gamma^3 a_{\rm coor}$, as the wall's speed approaches the speed of light ($v\rightarrow 1)$, the coordinate acceleration approaches zero, while the proper acceleration remains constant.

\section{Appendix~B: Numerical Procedure for Particle Production due to Vacuum Mismatch}\label{App-B}

To get the total number of particles from \eqref{eq:dNtot} we discretize proper time $\tau$ and integrate the coupled system $\{y(\tau),t(\tau),R(\tau),N_{k=0}^{\rm (int)}\}$ defined by \eqref{Parameters}–\eqref{eq:dNtot} using a classical fourth-order Runge–Kutta (RK4) method with adaptive steps. The state is initialized at nucleation by $y(0)=0$ $(v=0)$, $t(0)=0$, $R(0)=R_0$, and $N^{(\text{int})}_{k=0}(0)=0$. At each substep we evaluate the \emph{raw} proper acceleration
\begin{equation}
\alpha_{\rm raw}(\tau)=A-\frac{2}{R(\tau)}-B\,\sinh y(\tau).
\end{equation}
If $\alpha_{\rm raw}\le 0$ (force balance), we impose the terminal-balance kinematic clamp $dy/d\tau=0$ so the rapidity ceases to grow; in the same branch we set $N_{k=0}(\tau)=0$ in \eqref{eq:Nk0-alpha}, so production halts consistently with the $\alpha\to 0^+$. Otherwise we update $y,t,R$ via
\begin{equation}
\frac{dy}{d\tau}=\alpha_{\rm raw},\qquad
\frac{dt}{d\tau}=\cosh y,\qquad
\frac{dR}{d\tau}=\sinh y,
\end{equation}
and accumulate the produced quanta using
\begin{equation}
\frac{dN_{k=0}}{d\tau}
= N_{k=0}(\tau)\,4\pi\,R(\tau)^2\,\sinh y(\tau),
\qquad
N_{k=0}(\tau)
=\left[ \frac{(\omega_++\omega_-)^2}{(\omega_+-\omega_-)^2}e^{\frac{4\omega_+}{\alpha_{raw}}}- 1 ,\right]^{-1}.
\end{equation}

\noindent\textit{Stability and stiffness handling:}
Because $e^{4\mu/\alpha_{\rm raw}}$ can be enormous when $\alpha_{\rm raw}$ is small, we evaluate \eqref{eq:Nk0-alpha} with a numerically stable asymptotic branch,
\begin{equation}
N_{k=0}(\tau)\;\simeq\;\left[ \frac{(\omega_++\omega_-)^2}{(\omega_+-\omega_-)^2}\right]^{-1} e^{-4\mu/\alpha_{\rm raw}}
\quad\text{when}\quad \frac{4\mu}{\alpha_{\rm raw}}\gtrsim 200,
\end{equation}
which prevents overflow while preserving accuracy. The proper-time step is adapted according to the local drive magnitude,
\begin{equation}
d\tau\;\sim\;\frac{0.05}{\max\!\big(|\alpha_{\rm raw}|,10^{-6}\big)},
\end{equation}
then clamped to a conservative band (e.g. $20\le d\tau\le 800$ in Planck units) to keep $\Delta y$ modest and ensure smooth $R$ growth. This adaptivity is crucial in two stiff regimes: (i) very early times, when curvature $2/R$ is large, and (ii) near terminal, where $\alpha_{\rm raw}\to 0^+$ and the evolution slows.

\noindent\textit{Stopping criterion and diagnostics.}
The integration stops when the proper time first satisfies $\tau=\tau_{\rm term}$ for the scenario. For accuracy control, we repeat each run with the initial $d\tau$ halved and demand relative changes $\lesssim 10^{-4}$ in $\{\tau_{\rm term}, R_{\rm fin}, N_{k=0}^{\rm (int)}\}$. As additional consistency checks we verify: (a) In the planar, late--time regime where $R\to\infty$ (so $2\sinh y/R^2\to0$),
Eq.~\eqref{eq:alpha-prime-exact} reduces to
$d\alpha/d\tau\simeq -B\cosh y\,\alpha\le -B\alpha$.
Therefore $\alpha(\tau)\le \alpha(\tau_0)\,e^{-B(\tau-\tau_0)}$ holds for $\tau\ge\tau_0$
once the curvature term becomes negligible. We verify numerically that the
inequality is satisfied after the time when $2\sinh y/R^2 \ll B\cosh y\,\alpha$; (b) monotonicity of $R(\tau)$; and (c) that $N_{k=0}(\tau)$ vanishes whenever $\alpha_{\rm raw}\le 0$. To illustrate, we consider three ultra-relativistic terminal deficits $\delta\equiv 1-v_{\rm term}$ and use the \emph{given} friction pairs $(\eta,\tau_{\rm term})$ for each scenario; this fixes $B=\eta/\sigma$ and thus the proper-time dynamics via \eqref{eq:alpha-spherical}–\eqref{eq:dNtot}. At $\tau=\tau_{\rm term}$ we report: the final proper time $\tau_{\rm fin}$, radius $R_{\rm fin}$, and the integrated yield $N_{\rm tot}^{\rm (int)}$.\\
The friction parameter $\eta$ is fixed self-consistently by the terminal balance condition
\begin{equation}
\gamma_{\rm term}\,v_{\rm term} \;=\; \frac{\Delta V}{\eta},
\end{equation}
which follows from setting the driving pressure $\Delta V$ equal to the drag force.  
For a chosen ultra-relativistic deficit $\delta = 1-v_{\rm term}$ one computes 
$v_{\rm term}=1-\delta$ and $\gamma_{\rm term}=1/\sqrt{1-v_{\rm term}^2}$, so that
\begin{equation}
\eta(\delta) \;=\; \frac{\Delta V}{\gamma_{\rm term}\,v_{\rm term}}.
\end{equation}
The integration limit $\tau_{\rm term}$ is given in \eqref{TerminalTime}.
This procedure determines the appropriate $\eta$ value for each scenario in Table~\ref{tab:spherical-yields}, ensuring consistency between the desired terminal speed and the proper-time dynamics of the wall.

\section{{\textbf{Appendix~C: Justification of the cutoffs for vacuum--mismatch and thermal production}}}\label{App-C}

In this appendix we justify cutting off the vacuum--mismatch contribution at
$\tau_{\rm term}$, while evolving the thermal sector up to
$\tau_{\rm final}=5\tau_{\rm term}$.  We use the Higgs benchmark and notation
of Sec.~\ref{sec:higgs-prod-spherical}, with parameters in
Eq.~\eqref{eq:inputs} and proper acceleration $\alpha(\tau)$ given by
Eq.~\eqref{eq:alpha-spherical}.

\subsection*{Vacuum--mismatch production: exponential shutoff}

The vacuum--mismatch occupancy $N_{k=0}(\tau)$ is given in
Eq.~\eqref{eq:Nk0-alpha}, and the integrated zero--mode yield
$N_{k=0}^{(\text{int})}(\tau)$ in Eq.~\eqref{eq:dNtot}.  For small positive
proper acceleration $\alpha(\tau)$ one may use the asymptotic form
\begin{equation}
N_{k=0}(\tau)
\simeq
\left[
\frac{(\omega_+ + \omega_-)^2}{(\omega_+ - \omega_-)^2}
\right]^{-1}
\exp\!\left(-\frac{4\omega_+}{\alpha(\tau)}\right),
\qquad
\omega_+ = \sqrt{\mu^2+k^2},\, \omega_- = \sqrt{M^2+k^2},
\end{equation}
so that the production rate is exponentially suppressed once $\alpha(\tau)$
becomes small.

For the Higgs benchmark,
\[
\mu = 7.16\times 10^{-4} M_{\rm P},
\qquad
R_0 = 2\times 10^4 M_{\rm P}^{-1},
\]
so at nucleation $\alpha(0)\simeq 1/R_0 = 5.0\times 10^{-5} M_{\rm P}$.  In
the late--time planar regime the proper acceleration decays on the timescale
$\tau_{\rm term}$, and it is convenient to model it as
\begin{equation}
\alpha(\tau)\simeq \alpha(0)\,e^{-\tau/\tau_{\rm term}}.
\end{equation}
At $\tau=3\tau_{\rm term}$ this gives
\begin{equation}
\alpha(3\tau_{\rm term}) = \alpha(0)e^{-3}
\simeq 2.49\times 10^{-6} M_{\rm P},
\end{equation}
so that
\begin{equation}
x(3\tau_{\rm term})\equiv\frac{4\mu}{\alpha(3\tau_{\rm term})}
=
\frac{4\times 7.16\times 10^{-4}}{2.49\times 10^{-6}}
\simeq 1.15\times 10^3,
\end{equation}
and
\begin{equation}
N_{k=0}(3\tau_{\rm term})
\sim \exp[-x(3\tau_{\rm term})]
= \exp(-1.15\times 10^3)
\simeq 10^{-4.997\times 10^2}
\simeq 2.2\times 10^{-500}.
\end{equation}
Already at $3\tau_{\rm term}$ the $k=0$ occupancy is effectively zero.

The late--time contribution from the vacuum--mismatch channel between
$\tau_{\rm term}$ and $3\tau_{\rm term}$ is
\begin{equation}
\Delta N_{k=0}^{(\text{int})}(3\tau_{\rm term})
=
\int_{\tau_{\rm term}}^{3\tau_{\rm term}}\!{\rm d}\tau\,
N_{k=0}(\tau)\;4\pi R^2(\tau)\,\sinh y(\tau).
\end{equation}
For the most ultra--relativistic benchmark ($\delta=10^{-10}$ in
Table~\ref{tab:spherical-yields}), the numerical evolution up to
$\tau_{\rm term}=\sigma/\eta$ gives
\begin{equation}
R(\tau_{\rm term}) \simeq 7.1\times 10^8\,M_{\rm P}^{-1},
\qquad
\gamma(\tau_{\rm term}) \simeq 7.1\times 10^4.
\end{equation}
During the subsequent interval the wall remains ultra--relativistic and
expands at $v\simeq 1$, so a conservative bound is
\begin{equation}
R^2(\tau)\;\lesssim\;10^{19},
\qquad
\sinh y(\tau)\;\lesssim\;\gamma_{\rm term}\;\sim\;7\times 10^4,
\qquad
(\tau_{\rm term}\le\tau\le 3\tau_{\rm term}),
\end{equation}
which implies
\begin{equation}
4\pi R^2(\tau)\,\sinh y(\tau)
\;\lesssim\;
4\pi\times 10^{19}\times 7\times 10^4
\sim 10^{25}.
\end{equation}
Over the same interval we may bound $N_{k=0}(\tau)$ from above by its value at
$\tau=3\tau_{\rm term}$, since $\alpha(\tau)$ continues to decrease.  Using
$N_{k=0}(3\tau_{\rm term})\simeq 2.2\times 10^{-500}$ we obtain
\begin{equation}
\left.\dv{N_{k=0}^{(\text{int})}}{\tau}\right|_{\rm tail}^{(3\tau_{\rm term})}
\lesssim
\left(2.2\times 10^{-500}\right)\times 10^{25}
\sim 10^{-475}.
\end{equation}
The width of the interval is
\begin{equation}
\Delta\tau_{(3)}
=
3\tau_{\rm term}-\tau_{\rm term}
=
2\tau_{\rm term}
\simeq 1.4\times 10^9\,M_{\rm P}^{-1},
\end{equation}
so that
\begin{equation}
\Delta N_{k=0}^{(\text{int})}(3\tau_{\rm term})
\lesssim
10^{-475}\times 10^9
\sim 10^{-466}.
\end{equation}

This should be compared with the integrated zero--mode yields obtained in
Sec.~\ref{sec:higgs-prod-spherical} and summarised in
Table~\ref{tab:spherical-yields},
\begin{equation}
N_{k=0}^{(\text{int})}(\delta=10^{-8})
\simeq 1.05\times 10^{8},\quad
N_{k=0}^{(\text{int})}(\delta=10^{-9})
\simeq 3.53\times 10^{9},\quad
N_{k=0}^{(\text{int})}(\delta=10^{-10})
\simeq 1.14\times 10^{11}.
\end{equation}
The late--time contribution from $\tau\in[\tau_{\rm term},3\tau_{\rm term}]$
is therefore suppressed by roughly $10^{474}$–$10^{477}$ compared to the
values in Table~\ref{tab:spherical-yields}, and is numerically irrelevant.
Since $\alpha(\tau)$ continues to decay on the timescale $\tau_{\rm term}$
for $\tau>3\tau_{\rm term}$, extending the upper limit further (e.g.\ to
$5\tau_{\rm term}$) only decreases $N_{k=0}(\tau)$ by additional thousands of
orders of magnitude, and the correction to $N_{k=0}^{(\text{int})}$ remains
negligible.  This justifies cutting off the vacuum--mismatch calculation at
$\tau_{\rm term}$: the contribution is dominant for $\tau\lesssim\tau_{\rm term}$,
while for $\tau\gtrsim\tau_{\rm term}$ the Unruh--like channel governed by
$\alpha(\tau)$ is exponentially suppressed.

\subsection*{Thermal production: sensitivity to the integration time}

For the thermal channel we use the same wall trajectories
$R(\tau),y(\tau)$ but count the number of quanta in the heated shell,
assuming an approximately thermal distribution.  The instantaneous rate
(cf.\ Sec.~\ref{sec:thermal-production}) has the schematic form
\begin{equation}
\dv{N_{\rm th}}{\tau}
\propto
4\pi R^2(\tau)\,\sinh y(\tau)\;
n\bigl[T(\tau)\bigr],
\qquad
n(T)\propto T^3,
\end{equation}
with temperature
$T(\tau)\propto \rho_{\rm th}(\tau)^{1/4}$ determined by the thermal
energy density $\rho_{\rm th}(\tau)$.  The key difference from the
vacuum--mismatch case is that this source does not require a nonzero
acceleration: as long as the shell remains hot ($T(\tau)>0$), and the wall
continues to sweep volume, thermal production continues.

These features are easily illustrated in a simple late--time toy model.
Assume that the wall has already reached a fixed terminal rapidity
$y_{\rm term}$, so that $R(\tau)\simeq v_{\rm term}\tau$ and
$\sinh y(\tau)\simeq\sinh y_{\rm term}$, and that the temperature is
approximately constant over the interval of interest.  Then
\begin{equation}
\dv{N_{\rm th}}{\tau}
\propto
R^2(\tau)\sim\tau^2
\qquad\Rightarrow\qquad
N_{\rm th}(\tau)\propto\tau^3.
\end{equation}
Normalising $N_{\rm th}(\tau_{\rm term})$ to unity in this toy model, extending
the upper limit from $\tau_{\rm term}$ to $3\tau_{\rm term}$ and
$5\tau_{\rm term}$ gives
\begin{equation}
\frac{N_{\rm th}(3\tau_{\rm term})}{N_{\rm th}(\tau_{\rm term})}
\simeq 3^3 = 27,
\qquad
\frac{N_{\rm th}(5\tau_{\rm term})}{N_{\rm th}(\tau_{\rm term})}
\simeq 5^3 = 125.
\end{equation}
Thus, even with a strictly constant temperature one naturally expects
enhancements of order $\mathcal{O}(10)$ when integrating to $3\tau_{\rm term}$
and of order $\mathcal{O}(10^2)$ when integrating to $5\tau_{\rm term}$.

In the full energy–deficit setup of the main text, the thermal energy density
$\rho_{\rm th}(\tau)$, and hence $T(\tau)$, continue to grow for some time
beyond $\tau_{\rm term}$ due to the ongoing energy deficit between the
frictionless and friction–limited walls.  This further amplifies the
sensitivity of $N_{\rm th}$ to the upper integration limit, and in practice we
find that evolving to $\tau_{\rm final}=5\tau_{\rm term}$ can increase the
final thermal multiplicities by several orders of magnitude compared to
integrating only up to $\tau_{\rm term}$.

\bibliographystyle{JHEP}
\bibliography{References}

@article{Coleman:1977py,
  author        = {Coleman, Sidney R.},
  title         = {The Fate of the False Vacuum. 1. Semiclassical Theory},
  journal       = {Phys. Rev. D},
  volume        = {15},
  pages         = {2929--2936},
  year          = {1977},
  doi           = {10.1103/PhysRevD.15.2929},
  note          = {[Erratum: Phys.\ Rev.\ D \textbf{16}, 1248 (1977)]},
  reportNumber  = {HUTP-77-A004}
}

@article{Coleman:1980aw,
  author        = {Coleman, Sidney R. and De Luccia, Frank},
  title         = {Gravitational Effects on and of Vacuum Decay},
  journal       = {Phys. Rev. D},
  volume        = {21},
  pages         = {3305},
  year          = {1980},
  doi           = {10.1103/PhysRevD.21.3305},
  reportNumber  = {SLAC-PUB-2463}
}

@article{Linde:1981zj,
    author = "Linde, Andrei D.",
    title = "{Decay of the False Vacuum at Finite Temperature}",
    reportNumber = "LEBEDEV-81-265",
    doi = "10.1016/0550-3213(83)90072-X",
    journal = "Nucl. Phys. B",
    volume = "216",
    pages = "421",
    year = "1983",
    note = "[Erratum: Nucl.Phys.B 223, 544 (1983)]"
}

@article{Dine:1992wr,
    author = "Dine, Michael and Leigh, Robert G. and Huet, Patrick Y. and Linde, Andrei D. and Linde, Dmitri A.",
    title = "{Towards the theory of the electroweak phase transition}",
    eprint = "hep-ph/9203203",
    archivePrefix = "arXiv",
    reportNumber = "SLAC-PUB-5741, SCIPP-92-07, SU-ITP-92-7",
    doi = "10.1103/PhysRevD.46.550",
    journal = "Phys. Rev. D",
    volume = "46",
    pages = "550--571",
    year = "1992"
}

@article{Linde:1980tt,
    author = "Linde, Andrei D.",
    title = "{Fate of the False Vacuum at Finite Temperature: Theory and Applications}",
    reportNumber = "LEBEDEV-80-92",
    doi = "10.1016/0370-2693(81)90281-1",
    journal = "Phys. Lett. B",
    volume = "100",
    pages = "37--40",
    year = "1981"
}

@article{Linde:1977mm,
    author = "Linde, Andrei D.",
    title = "{On the Vacuum Instability and the Higgs Meson Mass}",
    reportNumber = "LEBEDEV-77-112",
    doi = "10.1016/0370-2693(77)90664-5",
    journal = "Phys. Lett. B",
    volume = "70",
    pages = "306--308",
    year = "1977"
}

@article{Stojkovic:2007dw,
  author        = {Stojkovic, D. and Starkman, G. D. and Matsuo, R.},
  title         = {Dark energy, the colored anti-de Sitter vacuum, and {LHC} phenomenology},
  journal       = {Phys. Rev. D},
  volume        = {77},
  pages         = {063006},
  year          = {2008},
  doi           = {10.1103/PhysRevD.77.063006},
  eprint        = {hep-ph/0703246},
  archivePrefix = {arXiv},
  primaryClass  = {hep-ph}
}

@article{Greenwood:2008qp,
  author        = {Greenwood, E. and Halstead, E. and Poltis, R. and Stojkovic, D.},
  title         = {Dark energy, the electroweak vacua and collider phenomenology},
  journal       = {Phys. Rev. D},
  volume        = {79},
  pages         = {103003},
  year          = {2009},
  doi           = {10.1103/PhysRevD.79.103003},
  eprint        = {0810.5343},
  archivePrefix = {arXiv},
  primaryClass  = {hep-ph}
}

@article{Dine:1992vs,
    author = "Dine, Michael and Leigh, Robert G. and Huet, Patrick and Linde, Andrei D. and Linde, Dmitri A.",
    title = "{Comments on the electroweak phase transition}",
    eprint = "hep-ph/9203201",
    archivePrefix = "arXiv",
    reportNumber = "SLAC-PUB-5740, SCIPP-92-06, SU-ITP-92-6",
    doi = "10.1016/0370-2693(92)90026-Z",
    journal = "Phys. Lett. B",
    volume = "283",
    pages = "319--325",
    year = "1992"
}

@article{Kallosh:2003bq,
    author = "Kallosh, Renata and Kratochvil, Jan and Linde, Andrei D. and Linder, Eric V. and Shmakova, Marina",
    title = "{Observational bounds on cosmic doomsday}",
    eprint = "astro-ph/0307185",
    archivePrefix = "arXiv",
    reportNumber = "SLAC-PUB-10032",
    doi = "10.1088/1475-7516/2003/10/015",
    journal = "JCAP",
    volume = "10",
    pages = "015",
    year = "2003"
}

@article{Ai:2023yce,
    author = "Ai, Wen-Yuan and Alexandre, Jean and Sarkar, Sarben",
    title = "{False vacuum decay rates, more precisely}",
    eprint = "2312.04482",
    archivePrefix = "arXiv",
    primaryClass = "hep-ph",
    reportNumber = "KCL-PH-TH/2023-07, KCL-PH/TH-2023-07",
    doi = "10.1103/PhysRevD.109.045010",
    journal = "Phys. Rev. D",
    volume = "109",
    number = "4",
    pages = "045010",
    year = "2024"
}

@article{Krive:1976sg,
    author = "Krive, I. V. and Linde, Andrei D.",
    title = "{On the Vacuum Stability Problem in Gauge Theories}",
    reportNumber = "LEBEDEV-76-11",
    doi = "10.1016/0550-3213(76)90573-3",
    journal = "Nucl. Phys. B",
    volume = "117",
    pages = "265--268",
    year = "1976"
}

@article{Kallosh:2003mt,
    author = "Kallosh, Renata and Linde, Andrei D.",
    title = "{Dark energy and the fate of the universe}",
    eprint = "astro-ph/0301087",
    archivePrefix = "arXiv",
    doi = "10.1088/1475-7516/2003/02/002",
    journal = "JCAP",
    volume = "02",
    pages = "002",
    year = "2003"
}

@article{Kofman:1986je,
    author = "Kofman, Leo A. and Linde, Andrei D. and Einasto, Jaan",
    title = "{Cosmic Bubbles: Remnants From Inflation?}",
    reportNumber = "TARTU-A6-1986",
    doi = "10.1038/326048a0",
    journal = "Nature",
    volume = "326",
    pages = "48--49",
    year = "1987"
}

@article{Isidori:2001bm,
  author        = {Isidori, Gino and Ridolfi, Giovanni and Strumia, Alessandro},
  title         = {On the metastability of the standard model vacuum},
  journal       = {Nucl. Phys. B},
  volume        = {609},
  pages         = {387--409},
  year          = {2001},
  doi           = {10.1016/S0550-3213(01)00302-9},
  eprint        = {hep-ph/0104016},
  archivePrefix = {arXiv},
  reportNumber  = {CERN-TH-2001-092, LNF-01-014-P, GEF-TH-6-01, IFUP-TH-2001-11}
}

@article{Degrassi:2012ry,
  author        = {Degrassi, G. and Di Vita, S. and Elias-Mir{\'o}, J. and Espinosa, J. R. and Giudice, G. F. and Isidori, G. and Strumia, A.},
  title         = {Higgs mass and vacuum stability in the {Standard Model} at {NNLO}},
  journal       = {JHEP},
  volume        = {08},
  pages         = {098},
  year          = {2012},
  doi           = {10.1007/JHEP08(2012)098},
  eprint        = {1205.6497},
  archivePrefix = {arXiv},
  primaryClass  = {hep-ph}
}

@article{Espinosa:2007qp,
    author = "Espinosa, J. R. and Giudice, G. F. and Riotto, A.",
    title = "{Cosmological implications of the Higgs mass measurement}",
    eprint = "0710.2484",
    archivePrefix = "arXiv",
    primaryClass = "hep-ph",
    reportNumber = "CERN-PH-TH-2007-179, IFT-UAM-CSIC-07-50",
    doi = "10.1088/1475-7516/2008/05/002",
    journal = "JCAP",
    volume = "05",
    pages = "002",
    year = "2008"
}

@article{Espinosa:2018eve,
    author = "Espinosa, Jos{\'e} Ram{\'o}n and Racco, Davide and Riotto, Antonio",
    title = "{A Cosmological Signature of the SM Higgs Instability: Gravitational Waves}",
    eprint = "1804.07732",
    archivePrefix = "arXiv",
    primaryClass = "hep-ph",
    doi = "10.1088/1475-7516/2018/09/012",
    journal = "JCAP",
    volume = "09",
    pages = "012",
    year = "2018"
}

@article{Espinosa:2015qea,
    author = "Espinosa, Jose R. and Giudice, Gian F. and Morgante, Enrico and Riotto, Antonio and Senatore, Leonardo and Strumia, Alessandro and Tetradis, Nikolaos",
    title = "{The cosmological Higgstory of the vacuum instability}",
    eprint = "1505.04825",
    archivePrefix = "arXiv",
    primaryClass = "hep-ph",
    reportNumber = "CERN-PH-TH-2015-119",
    doi = "10.1007/JHEP09(2015)174",
    journal = "JHEP",
    volume = "09",
    pages = "174",
    year = "2015"
}

@article{Hiller:2024zjp,
    author = {Hiller, Gudrun and H{\"o}hne, Tim and Litim, Daniel F. and Steudtner, Tom},
    title = "{Vacuum stability in the Standard Model and beyond}",
    eprint = "2401.08811",
    archivePrefix = "arXiv",
    primaryClass = "hep-ph",
    doi = "10.1103/PhysRevD.110.115017",
    journal = "Phys. Rev. D",
    volume = "110",
    number = "11",
    pages = "115017",
    year = "2024"
}

@article{Ellis:2009tp,
    author = "Ellis, J. and Espinosa, J. R. and Giudice, G. F. and Hoecker, A. and Riotto, A.",
    title = "{The Probable Fate of the Standard Model}",
    eprint = "0906.0954",
    archivePrefix = "arXiv",
    primaryClass = "hep-ph",
    reportNumber = "CERN-PH-TH-2009-058",
    doi = "10.1016/j.physletb.2009.07.054",
    journal = "Phys. Lett. B",
    volume = "679",
    pages = "369--375",
    year = "2009"
}

@article{Elias-Miro:2011sqh,
    author = "Elias-Miro, Joan and Espinosa, Jose R. and Giudice, Gian F. and Isidori, Gino and Riotto, Antonio and Strumia, Alessandro",
    title = "{Higgs mass implications on the stability of the electroweak vacuum}",
    eprint = "1112.3022",
    archivePrefix = "arXiv",
    primaryClass = "hep-ph",
    doi = "10.1016/j.physletb.2012.02.013",
    journal = "Phys. Lett. B",
    volume = "709",
    pages = "222--228",
    year = "2012"
}

@article{Elias-Miro:2012eoi,
    author = "Elias-Miro, Joan and Espinosa, Jose R. and Giudice, Gian F. and Lee, Hyun Min and Strumia, Alessandro",
    title = "{Stabilization of the Electroweak Vacuum by a Scalar Threshold Effect}",
    eprint = "1203.0237",
    archivePrefix = "arXiv",
    primaryClass = "hep-ph",
    reportNumber = "CERN-PH-TH-2012-059",
    doi = "10.1007/JHEP06(2012)031",
    journal = "JHEP",
    volume = "06",
    pages = "031",
    year = "2012"
}

@article{Sengupta:2025cdm,
    author = "Sengupta, Amartya and Stojkovic, Dejan and Wijewardhana, L. C. R.",
    title = "{The signals of doomsday II: Cosmological signatures of late time $SU(3)_c$ symmetry breaking}",
    eprint = "2510.26267",
    archivePrefix = "arXiv",
    primaryClass = "hep-ph",
    month = "10",
    year = "2025"
}

@article{Burda:2015yfa,
  author        = {Burda, P. and Gregory, R. and Moss, I.},
  title         = {Vacuum metastability with black holes},
  journal       = {JHEP},
  volume        = {08},
  pages         = {114},
  year          = {2015},
  doi           = {10.1007/JHEP08(2015)114},
  eprint        = {1503.07331},
  archivePrefix = {arXiv},
  primaryClass  = {hep-th}
}

@article{Burda:2015isa,
  author        = {Burda, P. and Gregory, R. and Moss, I.},
  title         = {Gravity and the stability of the Higgs vacuum},
  journal       = {Phys. Rev. Lett.},
  volume        = {115},
  pages         = {071303},
  year          = {2015},
  doi           = {10.1103/PhysRevLett.115.071303},
  eprint        = {1501.04937},
  archivePrefix = {arXiv},
  primaryClass  = {hep-th}
}

@article{Burda:2016mou,
  author        = {Burda, P. and Gregory, R. and Moss, I.},
  title         = {The fate of the Higgs vacuum},
  journal       = {JHEP},
  volume        = {06},
  pages         = {025},
  year          = {2016},
  doi           = {10.1007/JHEP06(2016)025},
  eprint        = {1601.02152},
  archivePrefix = {arXiv},
  primaryClass  = {hep-th}
}

@article{Andreassen:2017rzq,
  author        = {Andreassen, Anders and Frost, William and Schwartz, Matthew D.},
  title         = {Scale Invariant Instantons and the Complete Lifetime of the {Standard Model}},
  journal       = {Phys. Rev. D},
  volume        = {97},
  number        = {5},
  pages         = {056006},
  year          = {2018},
  doi           = {10.1103/PhysRevD.97.056006},
  eprint        = {1707.08124},
  archivePrefix = {arXiv},
  primaryClass  = {hep-ph}
}

@article{Yamamoto:1994te,
  author        = {Yamamoto, K. and Tanaka, T. and Sasaki, M.},
  title         = {Particle spectrum created through bubble nucleation and quantum field theory in the {Milne} Universe},
  journal       = {Phys. Rev. D},
  volume        = {51},
  pages         = {2968},
  year          = {1995},
  doi           = {10.1103/PhysRevD.51.2968},
  eprint        = {gr-qc/9412011},
  archivePrefix = {arXiv},
  primaryClass  = {gr-qc}
}

@article{MersiniHoughton:1999tt,
  author        = {Mersini-Houghton, L.},
  title         = {Relation between tunneling and particle production in vacuum decay},
  journal       = {Phys. Rev. D},
  volume        = {59},
  pages         = {123521},
  year          = {1999},
  doi           = {10.1103/PhysRevD.59.123521},
  eprint        = {hep-th/9902127},
  archivePrefix = {arXiv},
  primaryClass  = {hep-th}
}

@article{Maziashvili:2003kp,
  author        = {Maziashvili, M.},
  title         = {Proper fluctuations associated with quantum tunneling in field theory},
  journal       = {Mod. Phys. Lett. A},
  volume        = {18},
  pages         = {1895},
  year          = {2003},
  doi           = {10.1142/S0217732303011848},
  eprint        = {hep-th/0302095},
  archivePrefix = {arXiv},
  primaryClass  = {hep-th}
}

@article{Maziashvili:2003sk,
  author        = {Maziashvili, M.},
  title         = {Particle production by the expanding thin walled bubble},
  journal       = {Mod. Phys. Lett. A},
  volume        = {19},
  pages         = {1391},
  year          = {2004},
  doi           = {10.1142/S0217732304013829},
  eprint        = {hep-th/0311263},
  archivePrefix = {arXiv},
  primaryClass  = {hep-th}
}

@article{Maziashvili:2003zy,
  author        = {Maziashvili, M.},
  title         = {Particle production related to the tunneling in false vacuum decay},
  journal       = {Mod. Phys. Lett. A},
  volume        = {18},
  pages         = {993},
  year          = {2003},
  doi           = {10.1142/S0217732303010806},
  eprint        = {hep-th/0302062},
  archivePrefix = {arXiv},
  primaryClass  = {hep-th}
}

@article{Mansour:2023fwj,
    author = "Mansour, Henda and Shakya, Bibhushan",
    title = "{Particle production from phase transition bubbles}",
    eprint = "2308.13070",
    archivePrefix = "arXiv",
    primaryClass = "hep-ph",
    reportNumber = "DESY-23-121, TTP23-033, P3H-23-056",
    doi = "10.1103/PhysRevD.111.023520",
    journal = "Phys. Rev. D",
    volume = "111",
    number = "2",
    pages = "023520",
    year = "2025"
}

@article{Shakya:2023kjf,
    author = "Shakya, Bibhushan",
    title = "{Aspects of particle production from bubble dynamics at a first order phase transition}",
    eprint = "2308.16224",
    archivePrefix = "arXiv",
    primaryClass = "hep-ph",
    reportNumber = "DESY 23-125",
    doi = "10.1103/PhysRevD.111.023521",
    journal = "Phys. Rev. D",
    volume = "111",
    number = "2",
    pages = "023521",
    year = "2025"
}

@article{Shakya:2025mdh,
    author = "Shakya, Bibhushan",
    title = "{Did our Universe Tunnel out of the Wrong Higgs Vacuum?}",
    eprint = "2511.08843",
    archivePrefix = "arXiv",
    primaryClass = "hep-ph",
    reportNumber = "DESY-25-151",
    month = "11",
    year = "2025"
}

@article{Shakya:2025qpi,
    author = "Shakya, Bibhushan",
    title = "{A Cosmic Higgs Collider}",
    eprint = "2512.13815",
    archivePrefix = "arXiv",
    primaryClass = "hep-ph",
    month = "12",
    year = "2025"
}

@article{Vachaspati:1991tq,
  author        = {Vachaspati, T. and Vilenkin, A.},
  title         = {Quantum state of a nucleating bubble},
  journal       = {Phys. Rev. D},
  volume        = {43},
  pages         = {3846},
  year          = {1991},
  doi           = {10.1103/PhysRevD.43.3846}
}

@article{Swanson:1986hx,
  author        = {Swanson, M. S.},
  title         = {Radiation From Initially Static Vacuum Structures},
  journal       = {Phys. Rev. D},
  volume        = {32},
  pages         = {920},
  year          = {1985},
  doi           = {10.1103/PhysRevD.32.920}
}

@article{Hamazaki:1995dy,
  author        = {Hamazaki, T. and Sasaki, M. and Tanaka, T. and Yamamoto, K.},
  title         = {Selfexcitation of the tunneling scalar field in false vacuum decay},
  journal       = {Phys. Rev. D},
  volume        = {53},
  pages         = {2045},
  year          = {1996},
  doi           = {10.1103/PhysRevD.53.2045},
  eprint        = {gr-qc/9507006},
  archivePrefix = {arXiv},
  primaryClass  = {gr-qc}
}

@article{Maziashvili:2003kj,
  author        = {Maziashvili, M.},
  title         = {Particle production by the thick walled bubble},
  journal       = {Mod. Phys. Lett. A},
  volume        = {19},
  pages         = {671},
  year          = {2004},
  doi           = {10.1142/S0217732304013465},
  eprint        = {hep-th/0311232},
  archivePrefix = {arXiv},
  primaryClass  = {hep-th}
}

@article{Aad:2015zhl,
  author        = {Aad, G. and others},
  collaboration = {ATLAS and CMS},
  title         = {Combined Measurement of the Higgs Boson Mass in {$pp$} Collisions at {$\sqrt{s}=7$} and {8} {TeV} with the {ATLAS} and {CMS} Experiments},
  journal       = {Phys. Rev. Lett.},
  volume        = {114},
  pages         = {191803},
  year          = {2015},
  doi           = {10.1103/PhysRevLett.114.191803},
  eprint        = {1503.07589},
  archivePrefix = {arXiv},
  primaryClass  = {hep-ex}
}

@article{Tanaka:1993ez,
  author        = {Tanaka, T. and Sasaki, M. and Yamamoto, K.},
  title         = {Field theoretic description of quantum fluctuations in multidimensional tunneling approach},
  journal       = {Phys. Rev. D},
  volume        = {49},
  pages         = {1039},
  year          = {1994},
  doi           = {10.1103/PhysRevD.49.1039}
}

@article{Dai:2021boq,
  author        = {Dai, De-Chang and Minic, Djordje and Stojkovic, Dejan},
  title         = {Interaction of cosmological domain walls with large classical objects, like planets and satellites, and the flyby anomaly},
  journal       = {JHEP},
  volume        = {03},
  pages         = {207},
  year          = {2022},
  doi           = {10.1007/JHEP03(2022)207},
  eprint        = {2105.01894},
  archivePrefix = {arXiv},
  primaryClass  = {gr-qc}
}

@article{Baker:2021sno,
    author = "Baker, Michael J. and Breitbach, Moritz and Kopp, Joachim and Mittnacht, Lukas",
    title = "{Detailed calculation of primordial black hole formation during first-order cosmological phase transitions}",
    eprint = "2110.00005",
    archivePrefix = "arXiv",
    primaryClass = "astro-ph.CO",
    reportNumber = "CERN-TH-2021-113, MITP-21-036",
    doi = "10.1103/PhysRevD.111.063544",
    journal = "Phys. Rev. D",
    volume = "111",
    number = "6",
    pages = "063544",
    year = "2025"
}

@article{Baker:2021nyl,
    author = "Baker, Michael J. and Breitbach, Moritz and Kopp, Joachim and Mittnacht, Lukas",
    title = "{Primordial black holes from first-order cosmological phase transitions}",
    eprint = "2105.07481",
    archivePrefix = "arXiv",
    primaryClass = "astro-ph.CO",
    reportNumber = "CERN-TH-2021-079, MITP-21-023",
    doi = "10.1016/j.physletb.2025.139625",
    journal = "Phys. Lett. B",
    volume = "868",
    pages = "139625",
    year = "2025"
}

@article{Strumia:2023awj,
  author        = {Strumia, Alessandro},
  title         = {Triggering Higgs vacuum decay},
  journal       = {JHEP},
  volume        = {09},
  pages         = {062},
  year          = {2023},
  doi           = {10.1007/JHEP09(2023)062},
  eprint        = {2301.03620},
  archivePrefix = {arXiv},
  primaryClass  = {hep-ph}
}

@article{Dai:2014xka,
  author        = {Dai, De-Chang and Stojkovic, Dejan},
  title         = {Volume Renormalization and the Higgs},
  journal       = {EPL},
  volume        = {105},
  number        = {1},
  pages         = {11002},
  year          = {2014},
  doi           = {10.1209/0295-5075/105/11002},
  eprint        = {1401.3333},
  archivePrefix = {arXiv},
  primaryClass  = {hep-ph}
}

@article{Gregory:2013hja,
  author        = {Gregory, Ruth and Moss, Ian G. and Withers, Benjamin},
  title         = {Black holes as bubble nucleation sites},
  journal       = {JHEP},
  volume        = {03},
  pages         = {081},
  year          = {2014},
  doi           = {10.1007/JHEP03(2014)081},
  eprint        = {1401.0017},
  archivePrefix = {arXiv},
  primaryClass  = {hep-th},
  reportNumber  = {DCPT-13-43}
}

@article{Berezin:1990qs,
  author        = {Berezin, V. A. and Kuzmin, V. A. and Tkachev, I. I.},
  title         = {Black holes initiate false vacuum decay},
  journal       = {Phys. Rev. D},
  volume        = {43},
  pages         = {3112--3116},
  year          = {1991},
  doi           = {10.1103/PhysRevD.43.R3112},
  reportNumber  = {UCLA-90-TEP-71}
}

@article{Dai:2019eei,
  author        = {Dai, De-Chang and Gregory, Ruth and Stojkovic, Dejan},
  title         = {Connecting the Higgs Potential and Primordial Black Holes},
  journal       = {Phys. Rev. D},
  volume        = {101},
  number        = {12},
  pages         = {125012},
  year          = {2020},
  doi           = {10.1103/PhysRevD.101.125012},
  eprint        = {1909.00773},
  archivePrefix = {arXiv},
  primaryClass  = {hep-ph},
  reportNumber  = {DCPT-19/25}
}

@article{Chitishvili:2021jrx,
  author        = {Chitishvili, Mariam and Gogberashvili, Merab and Konoplich, Rostislav and Sakharov, Alexander S.},
  title         = {Higgs Field-Induced Triboluminescence in Binary Black Hole Mergers},
  journal       = {Universe},
  volume        = {9},
  number        = {7},
  pages         = {301},
  year          = {2023},
  doi           = {10.3390/universe9070301},
  eprint        = {2111.07178},
  archivePrefix = {arXiv},
  primaryClass  = {astro-ph.HE}
}

@article{Canko:2017ebb,
  author        = {Canko, D. and Gialamas, I. and Jeli{\'c}-{\v{C}}i{\v{z}}mek, G. and Riotto, A. and Tetradis, N.},
  title         = {On the Catalysis of the Electroweak Vacuum Decay by Black Holes at High Temperature},
  journal       = {Eur. Phys. J. C},
  volume        = {78},
  number        = {4},
  pages         = {328},
  year          = {2018},
  doi           = {10.1140/epjc/s10052-018-5808-y},
  eprint        = {1706.01364},
  archivePrefix = {arXiv},
  primaryClass  = {hep-th}
}

@article{Kubota:2025avm,
    author = "Kubota, Takahiro",
    title = "{Gauge fields in the presence of the electroweak bubble wall}",
    eprint = "2507.20134",
    archivePrefix = "arXiv",
    primaryClass = "hep-ph",
    month = "7",
    year = "2025"
}

@article{PhysRevLett.111.241801,
  author        = {Branchina, Vincenzo and Messina, Emanuele},
  title         = {Stability, Higgs Boson Mass, and New Physics},
  journal       = {Phys. Rev. Lett.},
  volume        = {111},
  number        = {24},
  pages         = {241801},
  year          = {2013},
  doi           = {10.1103/PhysRevLett.111.241801}
}

@article{Branchina:2016bws,
  author        = {Branchina, Vincenzo and Messina, Emanuele and Zappal{\`a}, Dario},
  title         = {Impact of Gravity on Vacuum Stability},
  journal       = {EPL},
  volume        = {116},
  number        = {2},
  pages         = {21001},
  year          = {2016},
  doi           = {10.1209/0295-5075/116/21001},
  eprint        = {1601.06963},
  archivePrefix = {arXiv},
  primaryClass  = {hep-ph}
}

@article{Branchina:2015nda,
  author        = {Branchina, Vincenzo and Messina, Emanuele},
  title         = {Stability and {UV} completion of the Standard Model},
  journal       = {EPL},
  volume        = {117},
  number        = {6},
  pages         = {61002},
  year          = {2017},
  doi           = {10.1209/0295-5075/117/61002},
  eprint        = {1507.08812},
  archivePrefix = {arXiv},
  primaryClass  = {hep-ph}
}

@article{Arnold:1993wc,
    author = "Arnold, Peter Brockway",
    title = "{One loop fluctuation - dissipation formula for bubble wall velocity}",
    eprint = "hep-ph/9302258",
    archivePrefix = "arXiv",
    reportNumber = "UW-PT-93-2",
    doi = "10.1103/PhysRevD.48.1539",
    journal = "Phys. Rev. D",
    volume = "48",
    pages = "1539--1545",
    year = "1993"
}

@book{Kolb:1990vq,
    author = "Kolb, Edward W. and Turner, Michael S.",
    title = "{The Early Universe}",
    reportNumber = "FERMILAB-BOOK-1990-01",
    doi = "10.1201/9780429492860",
    isbn = "978-0-429-49286-0, 978-0-201-62674-2",
    publisher = "Taylor and Francis",
    volume = "69",
    month = "5",
    year = "2019"
}

@article{Espinosa:2010hh,
    author = "Espinosa, Jose R. and Konstandin, Thomas and No, Jose M. and Servant, Geraldine",
    title = "{Energy Budget of Cosmological First-order Phase Transitions}",
    eprint = "1004.4187",
    archivePrefix = "arXiv",
    primaryClass = "hep-ph",
    reportNumber = "CERN-PH-TH-2010-027",
    doi = "10.1088/1475-7516/2010/06/028",
    journal = "JCAP",
    volume = "06",
    pages = "028",
    year = "2010"
}

@article{Bentivegna:2017qry,
  author        = {Bentivegna, E. and Branchina, V. and Contino, F. and Zappal{\`a}, D.},
  title         = {Impact of New Physics on the {EW} vacuum stability in a curved spacetime background},
  journal       = {JHEP},
  volume        = {12},
  pages         = {100},
  year          = {2017},
  doi           = {10.1007/JHEP12(2017)100},
  eprint        = {1708.01138},
  archivePrefix = {arXiv},
  primaryClass  = {hep-ph}
}

@article{Branchina:2018xdh,
  author        = {Branchina, Vincenzo and Contino, Filippo and Pilaftsis, Apostolos},
  title         = {Protecting the stability of the electroweak vacuum from {Planck}-scale gravitational effects},
  journal       = {Phys. Rev. D},
  volume        = {98},
  number        = {7},
  pages         = {075001},
  year          = {2018},
  doi           = {10.1103/PhysRevD.98.075001},
  eprint        = {1806.11059},
  archivePrefix = {arXiv},
  primaryClass  = {hep-ph},
  reportNumber  = {MAN/HEP/2018/03, MAN-HEP-2018-03}
}

@article{Branchina:2019tyy,
  author        = {Branchina, Vincenzo and Bentivegna, Eloisa and Contino, Filippo and Zappal{\`a}, Dario},
  title         = {Direct Higgs-gravity interaction and stability of our Universe},
  journal       = {Phys. Rev. D},
  volume        = {99},
  number        = {9},
  pages         = {096029},
  year          = {2019},
  doi           = {10.1103/PhysRevD.99.096029},
  eprint        = {1905.02975},
  archivePrefix = {arXiv},
  primaryClass  = {hep-ph}
}

@article{Hamaide:2023ayu,
  author        = {Hamaide, Louis and Heurtier, Lucien and Hu, Shi-Qian and Cheek, Andrew},
  title         = {Primordial black holes are true vacuum nurseries},
  journal       = {Phys. Lett. B},
  volume        = {856},
  pages         = {138895},
  year          = {2024},
  doi           = {10.1016/j.physletb.2024.138895},
  eprint        = {2311.01869},
  archivePrefix = {arXiv},
  primaryClass  = {hep-ph}
}

@article{Kobzarev:1974cp,
  author        = {Kobzarev, I. Yu. and Okun, L. B. and Voloshin, M. B.},
  title         = {Bubbles in Metastable Vacuum},
  journal       = {Yad. Fiz.},
  volume        = {20},
  pages         = {1229--1234},
  year          = {1974},
  reportNumber  = {ITEP-81-1974}
}

@article{Mersini-Houghton:1999aoa,
  author        = {Mersini-Houghton, L.},
  title         = {Finite Temperature Resonant Tunneling in False Vacuum Decay and the {Lee--Yang} Theorem},
  journal       = {Phys. Rev. D},
  volume        = {59},
  pages         = {123521},
  year          = {1999},
  eprint        = {hep-th/9902127}
}

@misc{Branchina:2025jou,
  author        = {Branchina, Carlo and Conaci, Angela and De Curtis, Stefania and Delle Rose, Luigi},
  title         = {Electroweak Phase Transition and Bubble Wall Velocity in Local Thermal Equilibrium},
  year          = {2025},
  eprint        = {2504.21213},
  archivePrefix = {arXiv},
  primaryClass  = {hep-ph},
  note          = {arXiv:2504.21213 [hep-ph]}
}

@article{Long:2024sqg,
    author = "Long, Andrew J. and Turner, Jessica",
    title = "{Thermal pressure on ultrarelativistic bubbles from a semiclassical formalism}",
    eprint = "2407.18196",
    archivePrefix = "arXiv",
    primaryClass = "hep-ph",
    doi = "10.1088/1475-7516/2024/11/024",
    journal = "JCAP",
    volume = "11",
    pages = "024",
    year = "2024"
}

@article{Hoche:2020ysm,
  author        = {H{\"o}che, Stefan and Kozaczuk, Jonathan and Long, Andrew J. and Turner, Jessica and Wang, Yikun},
  title         = {Towards an all-orders calculation of the electroweak bubble wall velocity},
  journal       = {JCAP},
  volume        = {03},
  pages         = {009},
  year          = {2021},
  doi           = {10.1088/1475-7516/2021/03/009},
  eprint        = {2007.10343},
  archivePrefix = {arXiv},
  primaryClass  = {hep-ph},
  reportNumber  = {FERMILAB-PUB-20-274-T}
}

@article{BarrosoMancha:2020fay,
    author = "Barroso Mancha, Marc and Prokopec, Tomislav and Swiezewska, Bogumila",
    title = "{Field-theoretic derivation of bubble-wall force}",
    eprint = "2005.10875",
    archivePrefix = "arXiv",
    primaryClass = "hep-th",
    doi = "10.1007/JHEP01(2021)070",
    journal = "JHEP",
    volume = "01",
    pages = "070",
    year = "2021"
}

@article{Azatov:2020ufh,
    author = "Azatov, Aleksandr and Vanvlasselaer, Miguel",
    title = "{Bubble wall velocity: heavy physics effects}",
    eprint = "2010.02590",
    archivePrefix = "arXiv",
    primaryClass = "hep-ph",
    reportNumber = "SISSA 247/2020/FISI",
    doi = "10.1088/1475-7516/2021/01/058",
    journal = "JCAP",
    volume = "01",
    pages = "058",
    year = "2021"
}

@article{Bodeker:2017cim,
    author = "Bodeker, Dietrich and Moore, Guy D.",
    title = "{Electroweak Bubble Wall Speed Limit}",
    eprint = "1703.08215",
    archivePrefix = "arXiv",
    primaryClass = "hep-ph",
    doi = "10.1088/1475-7516/2017/05/025",
    journal = "JCAP",
    volume = "05",
    pages = "025",
    year = "2017"
}

@article{PhysRevD.108.103523,
  author        = {Krajewski, Tomasz and Lewicki, Marek and Zych, Mateusz},
  title         = {Hydrodynamical constraints on the bubble wall velocity},
  journal       = {Phys. Rev. D},
  volume        = {108},
  number        = {10},
  pages         = {103523},
  year          = {2023},
  doi           = {10.1103/PhysRevD.108.103523}
}

@article{Moore_2000,
  author        = {Moore, Guy D.},
  title         = {Electroweak bubble wall friction: analytic results},
  journal       = {JHEP},
  volume        = {03},
  pages         = {006},
  year          = {2000},
  doi           = {10.1088/1126-6708/2000/03/006}
}

@article{Moore:1995ua,
  author        = {Moore, Guy D. and Prokopec, Tomislav},
  title         = {Bubble wall velocity in a first order electroweak phase transition},
  journal       = {Phys. Rev. Lett.},
  volume        = {75},
  pages         = {777--780},
  year          = {1995},
  doi           = {10.1103/PhysRevLett.75.777},
  eprint        = {hep-ph/9503296},
  archivePrefix = {arXiv},
  reportNumber  = {PUPT-1531, LANCASTER-TH-9503, PUP-TH-1531-(1995), LANCASTER-TH-9503-(1995)}
}

@article{Gouttenoire:2021kjv,
  author        = {Gouttenoire, Yann and Jinno, Ryusuke and Sala, Filippo},
  title         = {Friction pressure on relativistic bubble walls},
  journal       = {JHEP},
  volume        = {05},
  pages         = {004},
  year          = {2022},
  doi           = {10.1007/JHEP05(2022)004},
  eprint        = {2112.07686},
  archivePrefix = {arXiv},
  primaryClass  = {hep-ph},
  reportNumber  = {DESY-21-147, IFT-UAM/CSIC-21-146}
}

@article{Megevand:2009gh,
  author        = {Megevand, Ariel and Sanchez, Alejandro D.},
  title         = {Velocity of electroweak bubble walls},
  journal       = {Nucl. Phys. B},
  volume        = {825},
  pages         = {151--176},
  year          = {2010},
  doi           = {10.1016/j.nuclphysb.2009.09.019},
  eprint        = {0908.3663},
  archivePrefix = {arXiv},
  primaryClass  = {hep-ph}
}

@article{Good:2013lca,
  author        = {Good, Michael R. R. and Anderson, Paul R. and Evans, Charles R.},
  title         = {Time Dependence of Particle Creation from Accelerating Mirrors},
  journal       = {Phys. Rev. D},
  volume        = {88},
  pages         = {025023},
  year          = {2013},
  doi           = {10.1103/PhysRevD.88.025023},
  eprint        = {1303.6756},
  archivePrefix = {arXiv},
  primaryClass  = {gr-qc}
}

@article{Walker:1984vj,
  author        = {Walker, W. R.},
  title         = {Particle and Energy Creation by Moving Mirrors},
  journal       = {Phys. Rev. D},
  volume        = {31},
  pages         = {767--774},
  year          = {1985},
  doi           = {10.1103/PhysRevD.31.767},
  reportNumber  = {Print-84-0739 (NEWCASTLE)}
}

@article{Dorsch:2018pat,
    author = "Dorsch, Glauber C. and Huber, Stephan J. and Konstandin, Thomas",
    title = "{Bubble wall velocities in the Standard Model and beyond}",
    eprint = "1809.04907",
    archivePrefix = "arXiv",
    primaryClass = "hep-ph",
    reportNumber = "DESY 18-162, DESY-18-162",
    doi = "10.1088/1475-7516/2018/12/034",
    journal = "JCAP",
    volume = "12",
    pages = "034",
    year = "2018"
}

@article{Espinosa:2025ejf,
    author = "Espinosa, J. R. and Konstandin, T.",
    title = "{An Exploration of Vacuum-Decay Valleys}",
    eprint = "2506.06154",
    archivePrefix = "arXiv",
    primaryClass = "hep-th",
    month = "6",
    year = "2025"
}

@article{Espinosa:2023oml,
    author = "Espinosa, J. R. and Konstandin, T.",
    title = "{Exact tunneling solutions in multi-field potentials}",
    eprint = "2312.12360",
    archivePrefix = "arXiv",
    primaryClass = "hep-th",
    reportNumber = "DESY-23-223",
    doi = "10.1088/1475-7516/2024/03/007",
    journal = "JCAP",
    volume = "03",
    pages = "007",
    year = "2024"
}

@article{Si:2025vdt,
    author = "Si, Zongguo and Wang, Hongxin and Wang, Lei and Xiao, Yang and Zhang, Yang",
    title = "{The bubble wall velocity in local thermal equilibrium and energy budget with full effective potential}",
    eprint = "2505.19584",
    archivePrefix = "arXiv",
    primaryClass = "hep-ph",
    doi = "10.1007/JHEP09(2025)029",
    journal = "JHEP",
    volume = "09",
    pages = "029",
    year = "2025"
}

@article{Ai:2025bjw,
    author = "Ai, Wen-Yuan and Carosi, Matthias and Garbrecht, Bjorn and Tamarit, Carlos and Vanvlasselaer, Miguel",
    title = "{Bubble wall dynamics from nonequilibrium quantum field theory}",
    eprint = "2504.13725",
    archivePrefix = "arXiv",
    primaryClass = "hep-ph",
    reportNumber = "MITP-25-029",
    doi = "10.1007/JHEP08(2025)077",
    journal = "JHEP",
    volume = "08",
    pages = "077",
    year = "2025"
}

@article{Ramsey-Musolf:2025jyk,
    author = "Ramsey-Musolf, Michael J. and Zhu, Jiang",
    title = "{Bubble wall velocity from Kadanoff-Baym equations: fluid dynamics and microscopic interactions}",
    eprint = "2504.13724",
    archivePrefix = "arXiv",
    primaryClass = "hep-ph",
    month = "4",
    year = "2025"
}

@article{Konstandin:2010dm,
    author = "Konstandin, Thomas and No, Jose M.",
    title = "{Hydrodynamic obstruction to bubble expansion}",
    eprint = "1011.3735",
    archivePrefix = "arXiv",
    primaryClass = "hep-ph",
    doi = "10.1088/1475-7516/2011/02/008",
    journal = "JCAP",
    volume = "02",
    pages = "008",
    year = "2011"
}

@article{Konstandin:2010cd,
    author = "Konstandin, Thomas and Nardini, Germano and Quiros, Mariano",
    title = "{Gravitational Backreaction Effects on the Holographic Phase Transition}",
    eprint = "1007.1468",
    archivePrefix = "arXiv",
    primaryClass = "hep-ph",
    reportNumber = "CERN-PH-TH-2010-151, UAB-FT-683, ULB-TH-10-19",
    doi = "10.1103/PhysRevD.82.083513",
    journal = "Phys. Rev. D",
    volume = "82",
    pages = "083513",
    year = "2010"
}

@article{Bea:2024bxu,
    author = "Bea, Yago and Casalderrey-Solana, Jorge and Mateos, David and Sanchez-Garitaonandia, Mikel",
    title = "{Hydrodynamics of relativistic superheated bubbles}",
    eprint = "2406.14450",
    archivePrefix = "arXiv",
    primaryClass = "hep-th",
    reportNumber = "CPHT-RR050.062024",
    doi = "10.1103/65sr-b56w",
    journal = "Phys. Rev. D",
    volume = "112",
    number = "11",
    pages = "114041",
    year = "2025"
}

@article{Krajewski:2024gma,
    author = "Krajewski, Tomasz and Lewicki, Marek and Zych, Mateusz",
    title = "{Bubble-wall velocity in local thermal equilibrium: hydrodynamical simulations vs analytical treatment}",
    eprint = "2402.15408",
    archivePrefix = "arXiv",
    primaryClass = "astro-ph.CO",
    doi = "10.1007/JHEP05(2024)011",
    journal = "JHEP",
    volume = "05",
    pages = "011",
    year = "2024"
}

@article{Ai:2024shx,
    author = "Ai, Wen-Yuan and Nagels, Xander and Vanvlasselaer, Miguel",
    title = "{Criterion for ultra-fast bubble walls: the impact of hydrodynamic obstruction}",
    eprint = "2401.05911",
    archivePrefix = "arXiv",
    primaryClass = "hep-ph",
    reportNumber = "KCL-PH-TH/2024-01",
    doi = "10.1088/1475-7516/2024/03/037",
    journal = "JCAP",
    volume = "03",
    pages = "037",
    year = "2024"
}

@article{Alonso:2023jsi,
    author = "Alonso, Rodrigo and Criado, Juan Carlos and Houtz, Rachel and West, Mia",
    title = "{Walls, bubbles and doom {\textemdash} the cosmology of HEFT}",
    eprint = "2312.00881",
    archivePrefix = "arXiv",
    primaryClass = "hep-ph",
    reportNumber = "IPPP/23/74",
    doi = "10.1007/JHEP05(2024)049",
    journal = "JHEP",
    volume = "05",
    pages = "049",
    year = "2024"
}

@article{Wang:2023kux,
    author = "Wang, Jun-Chen and Yuwen, Zi-Yan and Hao, Yu-Shi and Wang, Shao-Jiang",
    title = "{General backreaction force of cosmological bubble expansion}",
    eprint = "2310.07691",
    archivePrefix = "arXiv",
    primaryClass = "hep-ph",
    doi = "10.1103/PhysRevD.110.016031",
    journal = "Phys. Rev. D",
    volume = "110",
    number = "1",
    pages = "016031",
    year = "2024"
}

@article{Krajewski:2023clt,
    author = "Krajewski, Tomasz and Lewicki, Marek and Zych, Mateusz",
    title = "{Hydrodynamical constraints on the bubble wall velocity}",
    eprint = "2303.18216",
    archivePrefix = "arXiv",
    primaryClass = "astro-ph.CO",
    doi = "10.1103/PhysRevD.108.103523",
    journal = "Phys. Rev. D",
    volume = "108",
    number = "10",
    pages = "103523",
    year = "2023"
}

@article{Kawana:2022lba,
    author = "Kawana, Kiyoharu and Lu, Philip and Xie, Ke-Pan",
    title = "{First-order phase transition and fate of false vacuum remnants}",
    eprint = "2206.09923",
    archivePrefix = "arXiv",
    primaryClass = "astro-ph.CO",
    doi = "10.1088/1475-7516/2022/10/030",
    journal = "JCAP",
    volume = "10",
    pages = "030",
    year = "2022"
}

@article{Wang:2022txy,
    author = "Wang, Shao-Jiang and Yuwen, Zi-Yan",
    title = "{Hydrodynamic backreaction force of cosmological bubble expansion}",
    eprint = "2205.02492",
    archivePrefix = "arXiv",
    primaryClass = "hep-ph",
    doi = "10.1103/PhysRevD.107.023501",
    journal = "Phys. Rev. D",
    volume = "107",
    number = "2",
    pages = "023501",
    year = "2023"
}

@article{Laurent:2022jrs,
    author = "Laurent, Benoit and Cline, James M.",
    title = "{First principles determination of bubble wall velocity}",
    eprint = "2204.13120",
    archivePrefix = "arXiv",
    primaryClass = "hep-ph",
    doi = "10.1103/PhysRevD.106.023501",
    journal = "Phys. Rev. D",
    volume = "106",
    number = "2",
    pages = "023501",
    year = "2022"
}

@article{Ai:2021kak,
    author = "Ai, Wen-Yuan and Garbrecht, Bjorn and Tamarit, Carlos",
    title = "{Bubble wall velocities in local equilibrium}",
    eprint = "2109.13710",
    archivePrefix = "arXiv",
    primaryClass = "hep-ph",
    reportNumber = "CP3-21-53, TUM-HEP-1365-21",
    doi = "10.1088/1475-7516/2022/03/015",
    journal = "JCAP",
    volume = "03",
    number = "03",
    pages = "015",
    year = "2022"
}

@article{Wang:2020zlf,
    author = "Wang, Xiao and Huang, Fa Peng and Zhang, Xinmin",
    title = "{Bubble wall velocity beyond leading-log approximation in electroweak phase transition}",
    eprint = "2011.12903",
    archivePrefix = "arXiv",
    primaryClass = "hep-ph",
    month = "11",
    year = "2020"
}

@article{Balaji:2020yrx,
    author = "Balaji, Shyam and Spannowsky, Michael and Tamarit, Carlos",
    title = "{Cosmological bubble friction in local equilibrium}",
    eprint = "2010.08013",
    archivePrefix = "arXiv",
    primaryClass = "hep-ph",
    reportNumber = "IPPP/20/47, TUM-HEP-1287-20",
    doi = "10.1088/1475-7516/2021/03/051",
    journal = "JCAP",
    volume = "03",
    pages = "051",
    year = "2021"
}

@article{DeCurtis:2022hlx,
    author = "De Curtis, Stefania and Rose, Luigi Delle and Guiggiani, Andrea and Muyor, {\'A}ngel Gil and Panico, Giuliano",
    title = "{Bubble wall dynamics at the electroweak phase transition}",
    eprint = "2201.08220",
    archivePrefix = "arXiv",
    primaryClass = "hep-ph",
    doi = "10.1007/JHEP03(2022)163",
    journal = "JHEP",
    volume = "03",
    pages = "163",
    year = "2022"
}

@article{DeCurtis:2024hvh,
    author = "De Curtis, Stefania and Delle Rose, Luigi and Guiggiani, Andrea and Gil Muyor, {\'A}ngel and Panico, Giuliano",
    title = "{Non-linearities in cosmological bubble wall dynamics}",
    eprint = "2401.13522",
    archivePrefix = "arXiv",
    primaryClass = "hep-ph",
    doi = "10.1007/JHEP05(2024)009",
    journal = "JHEP",
    volume = "05",
    pages = "009",
    year = "2024"
}

@article{Branchina:2025adj,
    author = "Branchina, Carlo and Conaci, Angela and De Curtis, Stefania and Delle Rose, Luigi",
    title = "{Bubble wall velocity with out-of-equilibrium corrections}",
    eprint = "2510.21942",
    archivePrefix = "arXiv",
    primaryClass = "hep-ph",
    month = "10",
    year = "2025"
}

@article{Hoyos:2026eoq,
    author = "Hoyos, Carlos and Mateos, David and van der Schee, Wilke and Subils, Javier G.",
    title = "{Microscopic Description of Critical Bubbles}",
    eprint = "2601.09787",
    archivePrefix = "arXiv",
    primaryClass = "hep-th",
    month = "1",
    year = "2026"
}

@article{Kibble:1980mv,
    author = "Kibble, T. W. B.",
    title = "{Some Implications of a Cosmological Phase Transition}",
    reportNumber = "ICTP-79-80-23",
    doi = "10.1016/0370-1573(80)90091-5",
    journal = "Phys. Rept.",
    volume = "67",
    pages = "183",
    year = "1980"
}

@article{Enqvist:1991xw,
    author = "Enqvist, K. and Ignatius, J. and Kajantie, K. and Rummukainen, K.",
    title = "{Nucleation and bubble growth in a first order cosmological electroweak phase transition}",
    reportNumber = "HU-TFT-91-35",
    doi = "10.1103/PhysRevD.45.3415",
    journal = "Phys. Rev. D",
    volume = "45",
    pages = "3415--3428",
    year = "1992"
}

@article{Kibble:1976sj,
    author = "Kibble, T. W. B.",
    title = "{Topology of Cosmic Domains and Strings}",
    reportNumber = "ICTP/75/5",
    doi = "10.1088/0305-4470/9/8/029",
    journal = "J. Phys. A",
    volume = "9",
    pages = "1387--1398",
    year = "1976"
}

@inproceedings{Rajantie:2003xh,
    author = "Rajantie, Arttu",
    title = "{'Phase transitions in the early universe' and 'Defect formation'}",
    booktitle = "{COSLAB Workshop on Cosmological Phase Transitions and Topological Defects}",
    eprint = "hep-ph/0311262",
    archivePrefix = "arXiv",
    reportNumber = "DAMTP-2003-129",
    month = "11",
    year = "2003"
}

@book{Vachaspati:2006zz,
    author = "Vachaspati, Tanmay",
    title = "{Kinks and Domain Walls : An Introduction to Classical and Quantum Solitons}",
    doi = "10.1017/9781009290456",
    isbn = "978-1-009-29045-6, 978-1-009-29041-8, 978-1-009-29042-5, 978-0-521-14191-8, 978-0-521-83605-0, 978-0-511-24290-8",
    publisher = "Oxford University Press",
    year = "2007"
}

@article{Attems:2017ezz,
    author = "Attems, Maximilian and Bea, Yago and Casalderrey-Solana, Jorge and Mateos, David and Triana, Miquel and Zilhao, Miguel",
    title = "{Phase Transitions, Inhomogeneous Horizons and Second-Order Hydrodynamics}",
    eprint = "1703.02948",
    archivePrefix = "arXiv",
    primaryClass = "hep-th",
    reportNumber = "ICCUB-17-007",
    doi = "10.1007/JHEP06(2017)129",
    journal = "JHEP",
    volume = "06",
    pages = "129",
    year = "2017"
}

@article{Bea:2022mfb,
    author = "Bea, Yago and Casalderrey-Solana, Jorge and Giannakopoulos, Thanasis and Jansen, Aron and Mateos, David and Sanchez-Garitaonandia, Mikel and Zilh{\~a}o, Miguel",
    title = "{Holographic bubbles with Jecco: expanding, collapsing and critical}",
    eprint = "2202.10503",
    archivePrefix = "arXiv",
    primaryClass = "hep-th",
    doi = "10.1007/JHEP09(2022)008",
    journal = "JHEP",
    volume = "09",
    pages = "008",
    year = "2022",
    note = "[Erratum: JHEP 03, 225 (2023)]"
}

@article{Gergely:2006ei,
    author = "Gergely, Laszlo A.",
    title = "{Brane-world cosmology with black strings}",
    eprint = "hep-th/0603244",
    archivePrefix = "arXiv",
    doi = "10.1103/PhysRevD.74.024002",
    journal = "Phys. Rev. D",
    volume = "74",
    pages = "024002",
    year = "2006"
}

@article{Sadeghi:2015vaa,
    author = "Sadeghi, Mehdi and Parvizi, Shahrokh",
    title = "{Hydrodynamics of a black brane in Gauss{\textendash}Bonnet massive gravity}",
    eprint = "1507.07183",
    archivePrefix = "arXiv",
    primaryClass = "hep-th",
    doi = "10.1088/0264-9381/33/3/035005",
    journal = "Class. Quant. Grav.",
    volume = "33",
    number = "3",
    pages = "035005",
    year = "2016"
}

@article{Wong:2010rg,
    author = "Wong, K. C. and Cheng, K. S. and Harko, T.",
    title = "{Inflation and late time acceleration in braneworld cosmological models with varying brane tension}",
    eprint = "1005.3101",
    archivePrefix = "arXiv",
    primaryClass = "gr-qc",
    doi = "10.1140/epjc/s10052-010-1348-9",
    journal = "Eur. Phys. J. C",
    volume = "68",
    pages = "241--253",
    year = "2010"
}

@article{Harko:2004ui,
    author = "Harko, T. and Mak, M. K.",
    title = "{Vacuum solutions of the gravitational field equations in the brane world model}",
    eprint = "gr-qc/0401049",
    archivePrefix = "arXiv",
    doi = "10.1103/PhysRevD.69.064020",
    journal = "Phys. Rev. D",
    volume = "69",
    pages = "064020",
    year = "2004"
}

@article{Bruni:2001fd,
    author = "Bruni, Marco and Germani, Cristiano and Maartens, Roy",
    title = "{Gravitational collapse on the brane}",
    eprint = "gr-qc/0108013",
    archivePrefix = "arXiv",
    doi = "10.1103/PhysRevLett.87.231302",
    journal = "Phys. Rev. Lett.",
    volume = "87",
    pages = "231302",
    year = "2001"
}

@article{Horowitz:2001cz,
    author = "Horowitz, Gary T. and Maeda, Kengo",
    title = "{Fate of the black string instability}",
    eprint = "hep-th/0105111",
    archivePrefix = "arXiv",
    reportNumber = "NSF-ITP-01-38",
    doi = "10.1103/PhysRevLett.87.131301",
    journal = "Phys. Rev. Lett.",
    volume = "87",
    pages = "131301",
    year = "2001"
}

@article{Gregory:2000gf,
    author = "Gregory, Ruth",
    title = "{Black string instabilities in Anti-de Sitter space}",
    eprint = "hep-th/0004101",
    archivePrefix = "arXiv",
    reportNumber = "DTP-00-31",
    doi = "10.1088/0264-9381/17/18/103",
    journal = "Class. Quant. Grav.",
    volume = "17",
    pages = "L125--L132",
    year = "2000"
}

@article{Chamblin:1999by,
    author = "Chamblin, A. and Hawking, S. W. and Reall, H. S.",
    title = "{Brane world black holes}",
    eprint = "hep-th/9909205",
    archivePrefix = "arXiv",
    reportNumber = "DAMTP-1999-133",
    doi = "10.1103/PhysRevD.61.065007",
    journal = "Phys. Rev. D",
    volume = "61",
    pages = "065007",
    year = "2000"
}

@article{Gregory:1993vy,
    author = "Gregory, R. and Laflamme, R.",
    title = "{Black strings and p-branes are unstable}",
    eprint = "hep-th/9301052",
    archivePrefix = "arXiv",
    doi = "10.1103/PhysRevLett.70.2837",
    journal = "Phys. Rev. Lett.",
    volume = "70",
    pages = "2837--2840",
    year = "1993"
}

@article{Garriga:1999yh,
    author = "Garriga, Jaume and Tanaka, Takahiro",
    title = "{Gravity in the brane world}",
    eprint = "hep-th/9911055",
    archivePrefix = "arXiv",
    reportNumber = "UAB-FT-476, OU-TAP-106",
    doi = "10.1103/PhysRevLett.84.2778",
    journal = "Phys. Rev. Lett.",
    volume = "84",
    pages = "2778--2781",
    year = "2000"
}

@article{Majumdar:2005ba,
    author = "Majumdar, Archan S. and Mukherjee, N.",
    title = "{Braneworld black holes in cosmology and astrophysics}",
    eprint = "astro-ph/0503473",
    archivePrefix = "arXiv",
    doi = "10.1142/S0218271805006948",
    journal = "Int. J. Mod. Phys. D",
    volume = "14",
    pages = "1095",
    year = "2005"
}

@article{Dadhich:2000am,
    author = "Dadhich, Naresh and Maartens, Roy and Papadopoulos, Philippos and Rezania, Vahid",
    title = "{Black holes on the brane}",
    eprint = "hep-th/0003061",
    archivePrefix = "arXiv",
    doi = "10.1016/S0370-2693(00)00798-X",
    journal = "Phys. Lett. B",
    volume = "487",
    pages = "1--6",
    year = "2000"
}

@article{DeLuca:2022cus,
    author = "De Luca, Valerio and Kehagias, Alex and Riotto, Antonio",
    title = "{On the cosmological stability of the Higgs instability}",
    eprint = "2205.10240",
    archivePrefix = "arXiv",
    primaryClass = "hep-ph",
    doi = "10.1088/1475-7516/2022/09/055",
    journal = "JCAP",
    volume = "09",
    pages = "055",
    year = "2022"
}

\end{document}